\documentclass[12pt]{article}

\usepackage[a4paper,margin=3cm]{geometry}

\usepackage{authblk}

\usepackage[parfill]{parskip}
\usepackage{setspace}

\usepackage[labelfont=bf,font={small,stretch=0.95}]{caption}

\usepackage[super,sort,compress,numbers]{natbib}

\usepackage{hyperref}
 
\usepackage{gensymb}
\usepackage{amsmath}
\usepackage{amssymb}

\usepackage{graphicx}
\usepackage{multirow}

\usepackage[version=4]{mhchem}

\usepackage{mathptmx}

\begin{document}

\title{\Large\bf Signatures of paracrystallinity in amorphous silicon}

\author[1]{Louise A. M. Rosset}
\author[2]{David A. Drabold}
\author[1]{Volker L. Deringer\thanks{volker.deringer@chem.ox.ac.uk}}

\affil[1]{Department of Chemistry, University of Oxford, Oxford, UK}
\affil[2]{Department of Physics and Astronomy, Ohio University, Athens, OH, USA}

\date{}

\maketitle

\setstretch{1.5}

{\bf
The structure of amorphous silicon (a-Si) has been studied for decades.
The two main theories are based on a continuous random network and on a `paracrystalline' model, respectively---the latter being defined as showing localized structural order resembling the crystalline state whilst retaining an overall amorphous network.
However, the extent of this local order has been unclear, and experimental data have led to conflicting interpretations.
Here we show that signatures of paracrystallinity in an otherwise disordered network are indeed compatible with the existing body of experimental observations for a-Si.
We use quantum-mechanically accurate, machine-learning-driven simulations to systematically sample the configurational space of quenched a-Si, thereby allowing us to elucidate the boundary between amorphization and crystallization. We analyze our dataset using structural and local-energy descriptors to show that paracrystalline models are consistent with experiments in both regards.
Our work provides a unified explanation for seemingly conflicting theories in one of the most widely studied amorphous networks.
}

\clearpage

Amorphous silicon (a-Si) is one of the most widely studied disordered network solids, \cite{Lewis-22-03, Treacy-12-02, Roorda-12-12, Deringer-21-01} owing in equal parts to fundamental interest and to its range of applications. In particular, a-Si has a larger band gap than its crystalline counterpart, which is useful for solar-cell heterojunctions and thin-film transistors, \cite{Powell-89-12, Fischer-23-03} while its low mechanical loss makes it a candidate next-generation interferometer mirror coating material in the detection of gravitational waves using the LIGO or VIRGO instruments. \cite{Birney-18-11, Adhikari-20-07}

A great challenge to understanding the `true' local structure of a-Si is that there are various preparation methods, including self-ion implantation, \cite{Acco-96-02} laser glazing, \cite{Kear-79-04} or evaporation, \cite{Roorda-91-08} and that the structure of the resulting films depends strongly on the way by which they were made. In particular, the density, \cite{Acco-96-02, Laaziri-95-11} coordination environments, \cite{Xie-13-08, Fortner-89-03} and the presence of voids \cite{Moss-69-11, Ohdomari-81-11} vary from one sample to the next. While some authors regard self-ion implanted a-Si as the highest quality a-Si, this must be understood to be only one example of the material, albeit superbly characterized.

From foundational work in the 1930s \cite{Rosenhain-27,Zachariasen-32-10} has emerged the currently most widely accepted model for the structure of a-Si, known as the continuous random network (CRN). The CRN model is characterized by minimal deviation from 4-fold coordination and complete absence of long-range structural order. Computations using bond-switching methods \cite{Wooten-85-04, Barkema-00-08} have helped to popularize the CRN model. While a-Si cannot be experimentally quenched from the melt in bulk form, \cite{Zallen-98-05} machine-learning- (ML-) based interatomic potentials \cite{Bartok-18-12} have recently enabled mol\-e\-c\-ular-dynamics (MD) simulations of quenching bulk a-Si at rates of 10$^{11}$ K s$^{-1}$ (Ref.~\citenum{Deringer-18-06}) and slower. \cite{Bernstein-19} Such rates are comparable to those used in laser quenching experiments. \cite{Baeri-82-12} 

Despite the simplicity of the CRN model, and the fact that it is now widely seen as the preferred way to describe a-Si, \cite{Lewis-22-03} this model is not without challenges. The main argument against the CRN model is that it fails to capture the degree of medium-range order seen in fluctuation electron microscopy (FEM) experiments on a-Si. \cite{Treacy-98-07} Instead, an alternative explanation consistent with FEM data has been proposed, \cite{Treacy-98-07, Voyles-01-11} known as the `paracrystalline' model. The latter is defined as a strained nanocrystal embedded in an amorphous CRN matrix, without sharp grain boundaries. \cite{Treacy-98-07} 
Such paracrystalline structures have recently been synthesized and experimentally and computationally characterized for the lighter homologue, elemental carbon. \cite{Tang-21-11}
However, the paracrystalline model for a-Si conflicts with other experimental data \cite{Treacy-12-02, Roorda-12-12} and for many only qualifies as a mixed-phase material. \cite{Wright-13, Lewis-22-03} For some authors, the answer lies in an intermediate network between disordered and ordered Si \cite{Wright-13} which would explain findings related to the low-energy excitations of a-Si, \cite{Liu-14-07} while others argue from calorimetric data that there exists a configurational gap between amorphous and crystalline networks. \cite{Kail-11, Drabold-11} 
In short, the long-standing `CRN vs paracrystalline' debate has not been fully resolved. \cite{Ziman-79}

In the present study, we probe the limit between amorphization and crystallization of simulated melt-quenched Si. We systematically sample the configurational space of a-Si with an accurate and efficient teacher--student ML approach \cite{Morrow-09-22} (Methods), which allows us to explore the existence of a middle ground between fully disordered and crystalline structures. Both system size and simulation time, unlocked by efficient ML methods, \cite{Deringer-18-06, Morrow-09-22} are key to a full exploration of competing phases and microstructures. The results lead us to propose a revised paracrystalline Si model that is consistent with high-quality structural and calorimetric experimental data. We quantify structural and energetic properties of a-Si models over the range from disorder to order, thereby allowing us to gain unprecedented insight into the co-existence of the CRN and paracrystalline phases.  In so doing, we show that realistic and experimentally compatible models of a-Si are able to accommodate a small but significant degree of local paracrystalline order, whilst overall remaining a disordered network.

\begin{figure*}
    \includegraphics[width=\linewidth]{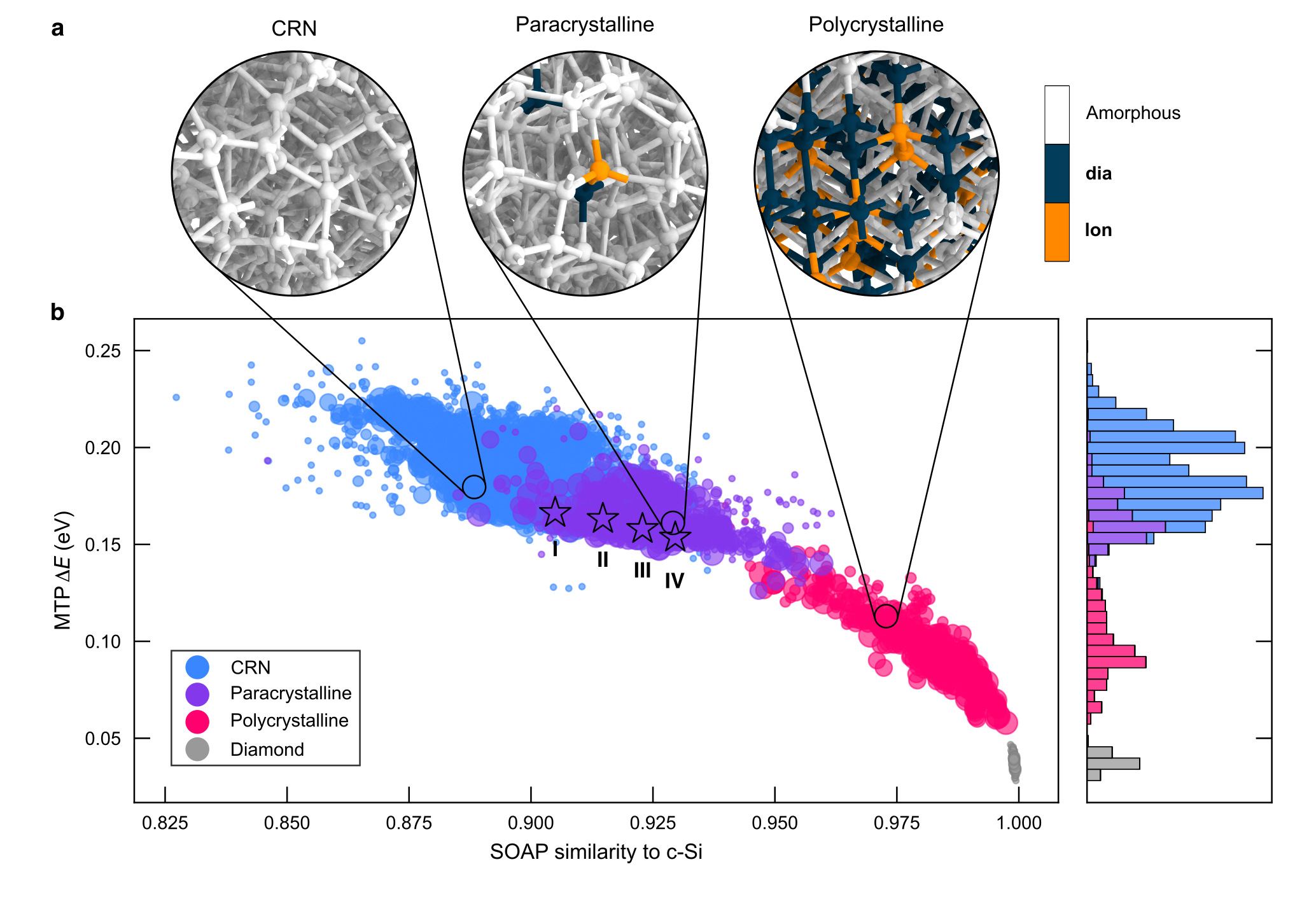}
    \caption{A comprehensive dataset of disordered Si structures. (a) Ball-and-stick rendering of representative structures from three categories, viz.\ continuous random network (CRN, left), paracrystalline (center), and polycrystalline (right). Polyhedral template matching was used to characterize atomic environments: blue indicates cubic-diamond-like environments ({\bf dia}), orange indicates hexagonal-diamond-like ({\bf lon}) ones, and white indicates atoms that do not fall within one of the defined categories (see Methods for details). (b) A map of similarity to diamond-type Si against the predicted excess energy (Methods). The marker sizes are proportional to the number of atoms in the respective structure. A stacked histogram of the energies is shown on the right, using the same vertical axis. The distributions in panel (b) indicate that the dataset spans structures from CRN- to diamond-like, encompassing a smooth range from disorder to gradual order.}
    \label{fig:DB}
\end{figure*}

We created a library of a-Si structural models in MD simulations with a systematically varied range of parameters. Specifically, we performed melt-quench simulations for four system sizes (64, 216, 512, and 1,000 atoms) with a uniform range of densities between 2.1 and 2.5 g cm$^{-3}$, over four quench rates of 10$^{13}$, 10$^{12}$, 10$^{11}$, and 10$^{10}$ K s$^{-1}$. To obtain a set of uncorrelated structures, we only take the final frame from each melt--quench simulation. This results in a dataset of 3,609 unique structures ($\approx 1.3$ million atoms). We note that in this part of the study, we focus on relatively small simulation cells on purpose; we will subsequently describe larger (100,000 atoms per cell) structural models.

Our dataset (Fig.~\ref{fig:DB}) contains structures ranging from highly disordered to very close to the crystalline form (c-Si). We characterize the dataset by plotting the computed excess energy, $\Delta E$ (relative to c-Si), against a measure for the similarity to the crystalline reference, where 1 is identical (Methods). We define structures as being either fully CRN-like, or paracrystalline, or polycrystalline using polyhedral template matching. \cite{Larsen-16-05} Some 64-atom structures  fully crystallized and formed strained diamond, shown in gray in Fig.~\ref{fig:DB}b.  

The fact that our dataset ranges almost smoothly from disorder to order (left$\,\rightarrow\,$right), both energetically and topologically, challenges the hypothesis of a `configurational energy gap' between c-Si and a-Si. \cite{Drabold-11} The paracrystalline structures populate the energetic middle ground between the CRN-like and polycrystalline configurations---which also challenges the initial theory of a {\em higher}-energy paracrystalline phase that could be annealed to yield a CRN. \cite{Treacy-98-07} While our dataset is relatively uniformly distributed, we observe a lower density of structures at the paracrystalline--polycrystalline transition, around 0.14 eV on the energy histogram in Fig.~\ref{fig:DB}. This corresponds to a deficit of structures with locally `crystal-like' environments between 15 and 40$\%$. These structures are likely absent from our dataset due to fast crystal-growth kinetics post nucleation, resulting in fewer structures with small crystalline grains. We note that some 64-atom a-Si structures (small markers) scatter widely in the plot of Fig.~\ref{fig:DB}, emphasizing that they sample a wider range of locally diverse environments, and thereby complement the larger structural models.

\begin{figure}
    \centering
    \includegraphics[width=0.6\linewidth]{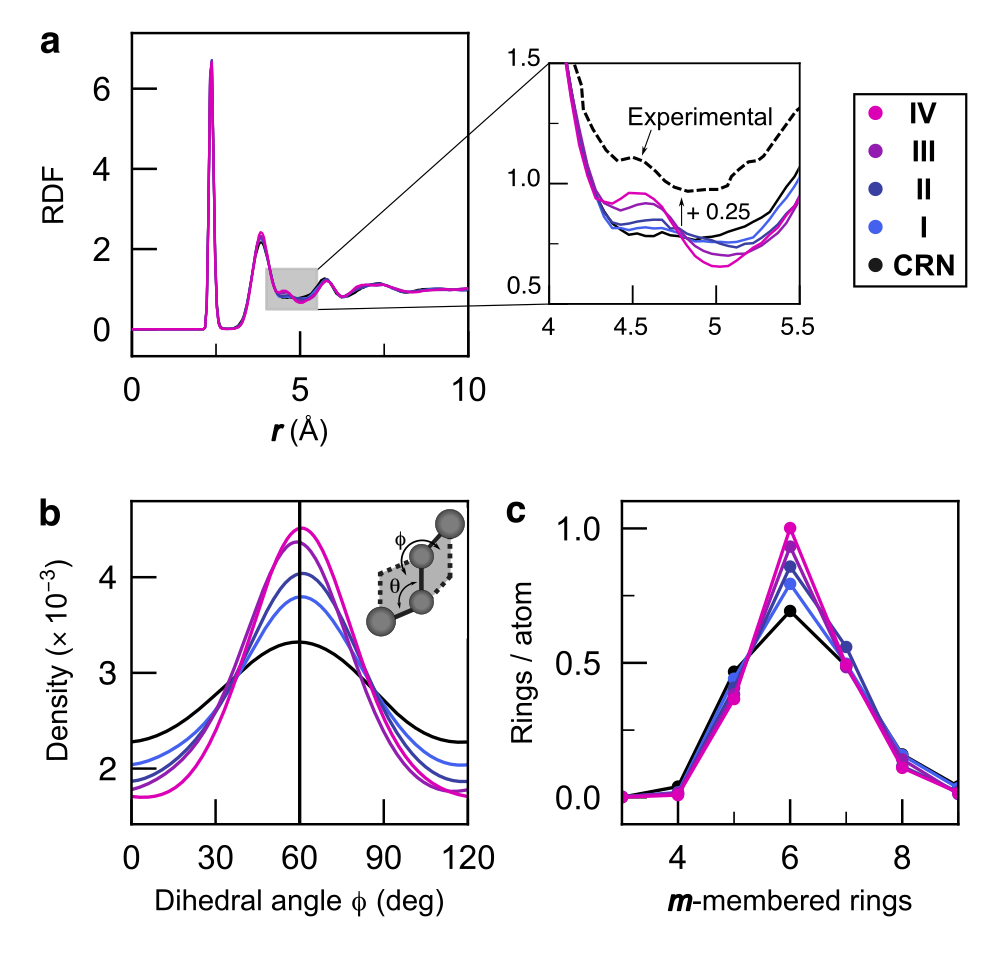}
    \caption{Characteristics of medium-range order for four structural models of increasing paracrystallinity ({\bf I}--{\bf IV}) as well as the CRN structure shown in Fig.~\ref{fig:DB}a. (a) Radial distribution function, and corresponding inset with experimental RDF from Ref.~\citenum{Laaziri-99-04}. (b) Dihedral angle distribution with a schematic indicating the definition of $\phi$. (c) Distribution of $m$-membered shortest-path rings.}
    \label{fig:MRO}
\end{figure}

While the paracrystalline category is intermediate between the CRN and polycrystalline ones, it shares significant topological and energetic overlap with the former. We select four paracrystalline structures of 1,000 atoms in the overlapping range, denoted {\bf I} to {\bf IV}, for more detailed analysis (Supplementary Fig.~8). These structures are increasingly paracrystalline, as reflected by their percentage of diamond-like environments of 0.2$\%$ ({\bf I}), 0.8$\%$ ({\bf II}), 2.4$\%$ ({\bf III}) and 4.5$\%$ ({\bf IV}). In Fig.~\ref{fig:MRO}, we use established indicators of short- and medium-range order to study these four structures. The radial distribution functions (RDFs) (Fig.~\ref{fig:MRO}a) are overall similar, with a well-defined valley between the first and second peak, indicating well-relaxed structures. The most relevant aspect in the context of paracrystallinity is the region between the second and third peaks, where experiments \cite{Laaziri-99-04, Laaziri-99-11, Fortner-89-03, Moss-70-08} showed a small but notable enhancement at about 4.5 \AA{}. Our series of models shows the gradual emergence of such a feature; the ratio between local maximum (at $\approx 4.5$ \AA{}) and local minimum (at $\approx 5.0$ \AA{}) is 1.08 for {\bf I} but 1.47 for {\bf IV}. Hence it is absent from the structure closest to CRN but replicated in the more paracrystalline structures. This feature has been attributed to a preferential orientation in the dihedral bond-angle distribution, \cite{Fortner-89-03, Laaziri-99-11} for which we show computed results in Fig.~\ref{fig:MRO}b. As paracrystallinity increases, the distribution sharpens while staying smooth---disagreeing with the claim that the RDF feature is only affected by the smoothness of the dihedral-angle distribution and not by its sharpness. \cite{Boissonade-74-10} Our results are qualitatively consistent with previous reports of paracrystalline signatures in the dihedral-angle distribution. \cite{Nakhmanson-01-05, Holmstrom-16-04} The shortest-path ring distribution (Fig.~\ref{fig:MRO}c) also mirrors the increasing degree of ordering from {\bf I} to {\bf IV}: 6-membered rings, characteristic of c-Si, become more abundant with paracrystallinity.

\begin{figure*}
    \centering
    \includegraphics[width=\linewidth]{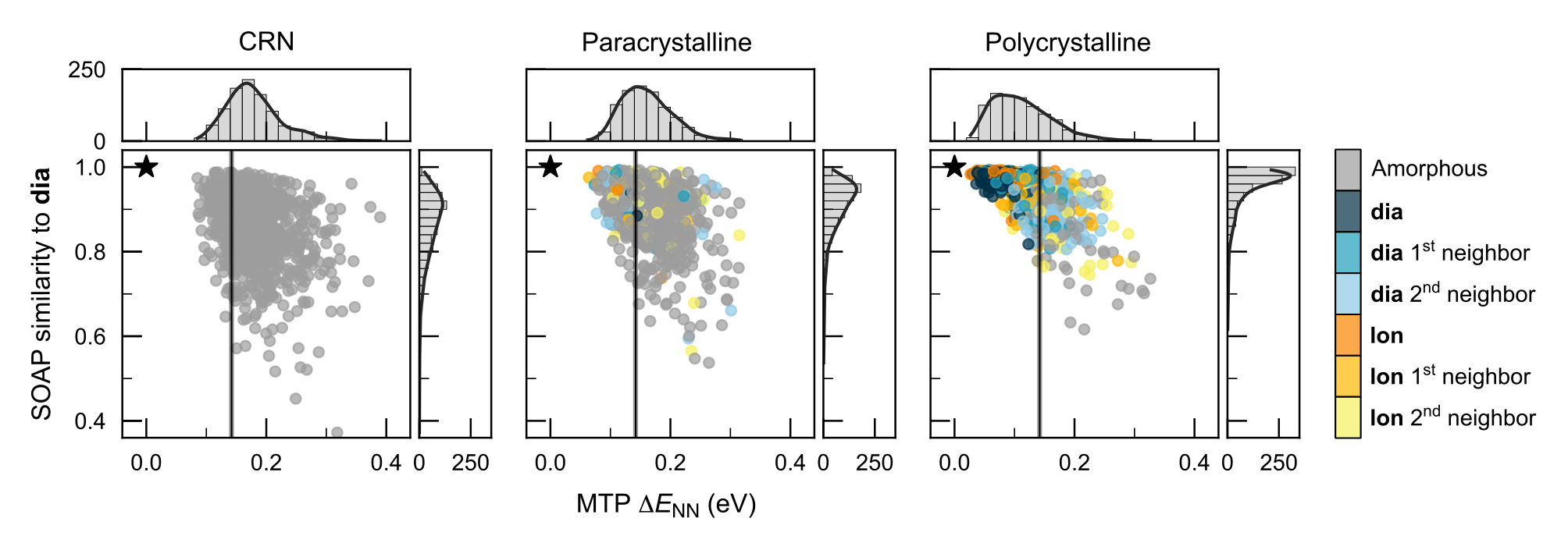}
    \caption{Energetics of disordered Si structures. Scatter plot of the ML-predicted atomic energy relative to cubic diamond-type Si ({\bf dia}) averaged over nearest neighbors against the atomistic SOAP similarity to {\bf dia}, colored by adaptive Common Neighbor Analysis.  A star indicates the ideal {\bf dia} environment. Histograms of the total distribution and kernel density estimates are shown for each axis. Vertical lines indicate the experimentally measured heat of crystallization, with gray shading corresponding to the standard deviation. \cite{Roorda-91-08}}
    \label{fig:deltaE}
\end{figure*}

Our analysis so far has established that the paracrystalline structures are structurally reasonable. The next step is to compare them directly with existing CRN models and to differentiate them from polycrystalline Si. In addition to structural information, it is important to consider energetic arguments. In Fig.~\ref{fig:deltaE}, we therefore focus on the local-energy fingerprints which can be derived from machine-learned atomic energies (Methods). We have shown previously that such an approach can help to map out the space of disorder and local order in monolayer amorphous carbon, \cite{ElMachachi-22-10} for which the distinction between CRN and (para-) crystallite descriptions has also been explored. \cite{ElMachachi-22-10, Toh-20-01} The present analysis in Fig.~\ref{fig:deltaE} hence takes us conceptually from a canonical disordered 2D system, amorphous graphene, to the canonical 3D case, which is a-Si.

For each of the three representative structures shown in Fig.~\ref{fig:DB}a, we represent the individual atomic environments therein as circles in Fig.~\ref{fig:deltaE}. We plot their computed excess energy, $\Delta E$ (relative to c-Si), averaged over their nearest neighbors, against a structural metric that quantifies how similar a given atom is to cubic-diamond-like Si (SOAP; Methods). We color-code the points based on Common Neighbor Analysis (CNA; Methods). 

Figure \ref{fig:deltaE} allows us to characterize the three fundamental forms that disordered silicon can take. The CRN structure shows only amorphous-like atomic environments, as expected. The energy histogram (horizontal axis) and SOAP similarity histogram (vertical axis) both show a single peak with a long tail. For the paracrystalline structure, some {\bf dia} and {\bf lon} environments are identified by CNA, but the majority of atomic environments are still amorphous-like. These diamond environments are far from the ideal diamond environment (star); they are not clustered together but distributed among the amorphous environments. The tails in both histograms are shorter, indicating that the amorphous environments in the paracrystalline structure do not suffer from additional strain from the presence of the localized diamond environments. Finally, for the polycrystalline structure, diamond-like environments are distinct from amorphous ones in both energy and structure. {\bf dia} and {\bf lon} environments are much closer to the ideal diamond environment than those in the paracrystalline structure are. The energy and SOAP histograms are characterized by two contributions, one from diamond-like and one from amorphous-like environments. Thus, the paracrystalline Si structures are comparable to the CRN ones, and can be delineated from the polycrystalline structures. We can ascertain that they are disordered, with localized crystal-like signatures.

The experimentally measured heat of crystallization, $\Delta H=0.142$ eV/atom, \cite{Roorda-91-08} is plotted alongside our ML local atomic energies in Fig.~\ref{fig:deltaE}. The paracrystalline structure agrees very well with these calorimetric data, where the CRN model is more energetic and the polycrystalline model is too stable compared to $\Delta H$. The paracrystalline structure also provides better agreement to $\Delta H$ than previous CRN models in the literature. \cite{Bernstein-19} 

\begin{figure*}
    \centering
    \includegraphics[width=\linewidth]{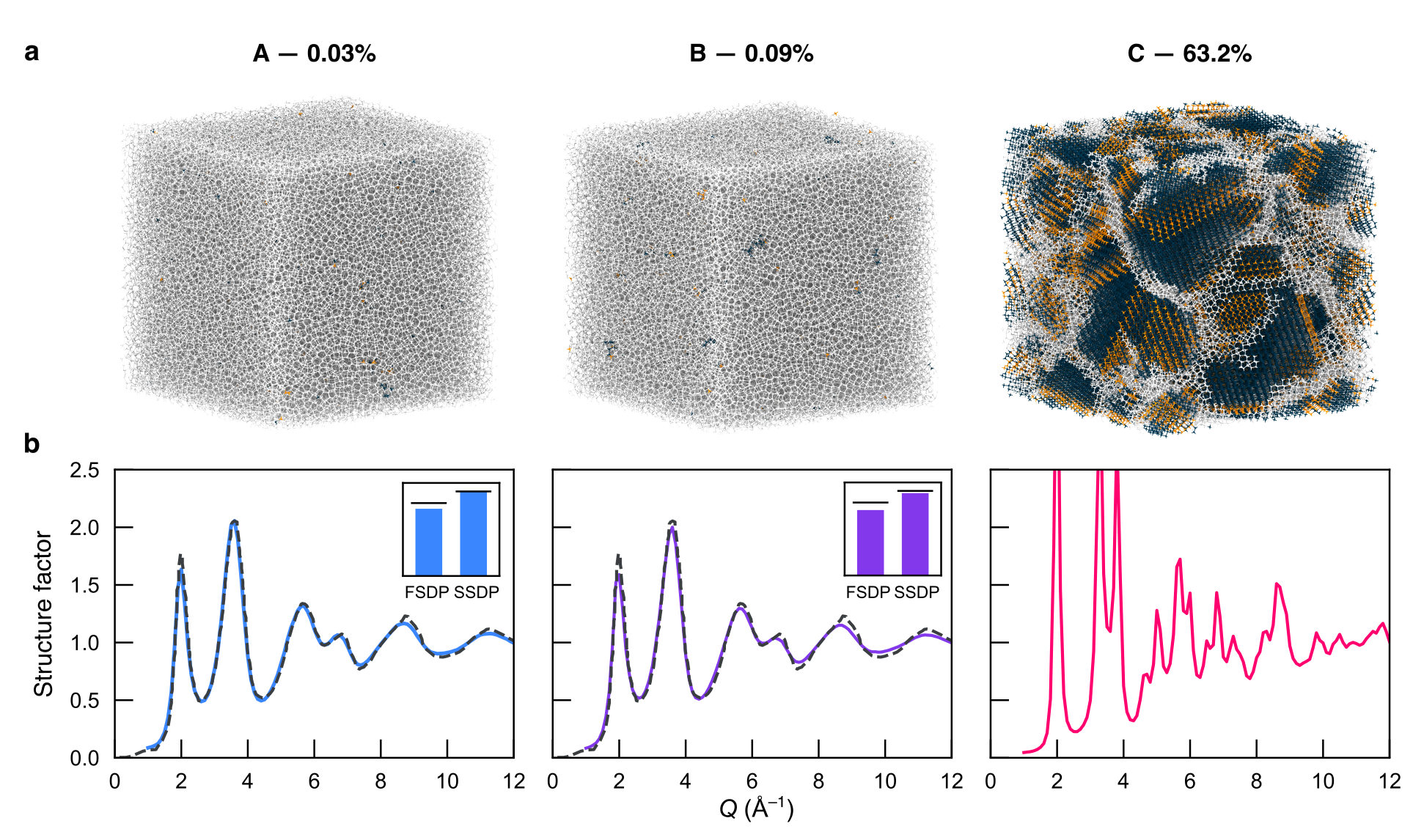}
    \caption{Three a-Si structural models of 100,000 atoms with increasing paracrystallinity. The first model, labeled \textbf{A}, is taken from Ref.~\citenum{Deringer-21-01}, while the other two were generated as part of the present study by melt-quenching at 10$^{11}$ K/s (\textbf{B}) and 10$^{10}$ K/s (\textbf{C}), respectively. (a) Structure visualizations color-coded by PTM as in Fig.~\ref{fig:DB}. (b) Computed structure factor for each structure, using the \textsc{DebyeCalculator} package. \cite{Johansen-24-02} Black dashed lines indicate the experimental data from Ref.~\citenum{Laaziri-99-04}. Insets show the agreement of the predicted first and second sharp diffraction peaks (bars) with experimental data from Ref.~\citenum{Laaziri-99-04} (black lines).}
    \label{fig:100k}
\end{figure*}

While our dataset provides valuable insight into the `middle ground' between fully disordered and crystalline silicon, the fact that we have used relatively small system sizes limits the comparability to experimental data. We therefore turn to a study on more realistic length scales, viz.\ $> 10$ nm, by preparing para- and polycrystalline models on that length scale using MD simulations, yielding models with 0.09$\%$ and 62.3$\%$ of diamond-like environments respectively. We compare against the structural model of Ref.~\citenum{Deringer-21-01} which had been created in simulations of the same type but driven by the teacher model, Si-GAP-18, and has 0.03$\%$ of diamond-like environments. These structures are shown side-by-side in Fig.~\ref{fig:100k}. The structure factor, $S(q)$, for each model is plotted together with high-quality experimental data from Ref.~\citenum{Laaziri-99-04}. The latter are well reproduced by the model with the lowest paracrystallinity \cite{Deringer-21-01}---but also by a more paracrystalline model, which is just as compatible with the experimental data. This implies that localized order can retain model agreement with experimental data, but only a small degree of crystallinity is beneficial as shown by the polycrystalline model.

In conclusion, we have systematically sampled the configurational space of Si, from fully disordered CRN-like networks to the diamond-type crystal, with extensive ML-driven atomistic simulations. Our results point toward a revised model for paracrystalline Si, at the limit between amorphization and crystallization, characterized by localized diamond-like neighborhoods that affect medium-range order. 
Paracrystalline structures show better agreement with high-quality experimental data for medium-range structural order and energetics than do previously proposed models. We note that while high-quality experiments are typically carried out on ion-implanted a-Si samples, laser-glazed a-Si is much closer to the melt-quenched samples generated by MD simulations. Further experimental work on laser-glazed a-Si could provide a closer basis for comparison, informing future theoretical and computational studies.

Our work opens important new avenues of exploration. As our dataset spans an essentially complete range of disorder, it is of interest to explore emergent phenomena unique to disordered matter such as the process of photodegredation known as the Staebler--Wronski effect, \cite{Staebler-77-08, Fedders-92-03} which could be investigated by computationally hydrogenating the structures in our dataset.
Two-level tunneling systems (TLSs), described as the tunneling between neighboring minima in the potential-energy landscape of amorphous materials, are also of fundamental interest for a-Si as they offer an explanation for low-energy excitations found at low temperatures. \cite{Fedders-96-02} A proposed origin for TLSs is nanoscale heterogeneity in the microstructure, taking the form of local order \cite{Liu-14-07}---such heterogeneity has been out of range for direct quantum-mechanical simulations, but is accessible using ML. \cite{Erhard-24-03} Systematically searching for perturbations that result in pairs of nearly identical amorphous configurations along the dataset's range from disorder to order could determine what extent of structural disorder in the network is required to observe tunneling. \cite{Guttman-86-01} Our work provides a high-quality dataset for further exploration of a-Si, and more widely exemplifies the role of ML in understanding fundamental phenomena in disordered materials.

\clearpage

\section*{Methods}

{\bf Teacher--student potentials.} The simulations in this work are based on a teacher--student machine-learning approach: \cite{Morrow-09-22} distilling an accurate, but comparably slow `teacher' ML potential (Si-GAP-18; Ref.~\citenum{Bartok-18-12}) into a faster `student' model, here using the Moment Tensor Potential (MTP) approach. \cite{Shapeev-16-09} We use the M$_{16}''$ model of Ref.~\citenum{Morrow-09-22}, which provides accuracy approaching that of Si-GAP-18 within the target domain (a-Si), whilst being $>100$ times faster. The teacher model has been extensively validated against experimental data for ambient and high-pressure a-Si; \cite{Deringer-18-06, Deringer-21-01} the student model enabled recent studies of coordination defects. \cite{Morrow-24-05} 

{\bf Structural analysis.} We classify structures as being either fully CRN-like, or paracrystalline, or polycrystalline using polyhedral template matching of atomic environments (PTM; RMSD cutoff of 0.1; Ref.~\citenum{Larsen-16-05}) as implemented in \textsc{OVITO}, \cite{Stukowski-10-01} with the following criteria: (i) if a structure contains no locally `crystal-like' atom, it is classified as fully CRN-like (blue in Fig.~\ref{fig:DB}); if it contains (ii) fewer or (iii) more than 15$\%$ of locally `crystal-like' atoms, it is classified conversely as paracrystalline (purple) or polycrystalline (magenta). The `polycrystalline' category is diverse, from large crystalline grains in an amorphous matrix to diamond structures with stacking faults.

For the analysis of local atomic environments, we employ two complementary techniques. First, we use the Smooth Overlap of Atomic Positions (SOAP) kernel \cite{Bartok-13-05} to quantify the similarity to the ideal diamond-type structure on a scale from 0 (dissimilar) to 1 (identical to within the cutoff radius), as done in previous work on a-Si. \cite{Bernstein-19, Morrow-09-22}
Second, we use Common Neighbor Analysis (CNA) \cite{Honeycutt-87-09} to identify the similarity to prototype structure types (specifically, \textbf{dia} and \textbf{lon}), as detailed in Ref.~\citenum{Maras-16-08}, and similar to Ref.~\citenum{Tang-21-11}.

{\bf Energetic analysis.} In many ML-based interatomic potentials, including the MTP framework, the total energy of a cell is constructed as the sum of the ML-learned individual atomic energies, \cite{Behler-07-04, Bartok-10-04} viz.\ $E=\sum_i E_i$. The distribution of such atomic energies has been shown to reveal the local stability of atoms in systems ranging from a-Si \cite{Bernstein-19} to superionic conductors. \cite{Wang-23-02} We further take the local atomic energies averaged over nearest neighbors, similar to our study of amorphous graphene, \cite{ElMachachi-22-10} here within a cutoff of 2.85 \AA{}. \cite{Bernstein-19} 

\clearpage

\section*{Acknowledgements}

We thank S.~R.~Elliott and M.~Wilson for helpful comments on the manuscript.
This work was supported through a UK Research and Innovation Frontier Research grant [grant number EP/X016188/1]. We are grateful for computational support from the UK national high performance computing service, ARCHER2, for which access was obtained via the UKCP consortium and funded by EPSRC grant ref EP/X035891/1.

\section*{Data availability}

Data supporting this study will be made openly available upon journal publication.

\section*{References}
\vspace{2mm}

\setstretch{1}


\begin{thebibliography}{10}
\expandafter\ifx\csname url\endcsname\relax
  \def\url#1{\texttt{#1}}\fi
\expandafter\ifx\csname urlprefix\endcsname\relax\def\urlprefix{URL }\fi
\providecommand{\bibinfo}[2]{#2}
\providecommand{\eprint}[2][]{\url{#2}}

\bibitem{Lewis-22-03}
\bibinfo{author}{Lewis, L.~J.}
\newblock \bibinfo{title}{Fifty years of amorphous silicon models : The end of the story?}
\newblock \emph{\bibinfo{journal}{J. Non-Cryst. Solids}} \textbf{\bibinfo{volume}{580}}, \bibinfo{pages}{121383} (\bibinfo{year}{2022}).

\bibitem{Treacy-12-02}
\bibinfo{author}{Treacy, M. M.~J.} \& \bibinfo{author}{Borisenko, K.~B.}
\newblock \bibinfo{title}{The {{Local Structure}} of {{Amorphous Silicon}}}.
\newblock \emph{\bibinfo{journal}{Science}} \textbf{\bibinfo{volume}{335}}, \bibinfo{pages}{950--953} (\bibinfo{year}{2012}).

\bibitem{Roorda-12-12}
\bibinfo{author}{Roorda, S.} \& \bibinfo{author}{Lewis, L.~J.}
\newblock \bibinfo{title}{Comment on ``{{The Local Structure}} of {{Amorphous Silicon}}''}.
\newblock \emph{\bibinfo{journal}{Science}} \textbf{\bibinfo{volume}{338}}, \bibinfo{pages}{1539--1539} (\bibinfo{year}{2012}).

\bibitem{Deringer-21-01}
\bibinfo{author}{Deringer, V.~L.} \emph{et~al.}
\newblock \bibinfo{title}{Origins of structural and electronic transitions in disordered silicon}.
\newblock \emph{\bibinfo{journal}{Nature}} \textbf{\bibinfo{volume}{589}}, \bibinfo{pages}{59--64} (\bibinfo{year}{2021}).

\bibitem{Powell-89-12}
\bibinfo{author}{Powell, M.}
\newblock \bibinfo{title}{The physics of amorphous-silicon thin-film transistors}.
\newblock \emph{\bibinfo{journal}{IEEE Trans. Electron Devices}} \textbf{\bibinfo{volume}{36}}, \bibinfo{pages}{2753--2763} (\bibinfo{year}{1989}).

\bibitem{Fischer-23-03}
\bibinfo{author}{Fischer, B.} \emph{et~al.}
\newblock \bibinfo{title}{The {{Microstructure}} of {{Underdense Hydrogenated Amorphous Silicon}} and its {{Application}} to {{Silicon Heterojunction Solar Cells}}}.
\newblock \emph{\bibinfo{journal}{Solar RRL}} \textbf{\bibinfo{volume}{7}}, \bibinfo{pages}{2300103} (\bibinfo{year}{2023}).

\bibitem{Birney-18-11}
\bibinfo{author}{Birney, R.} \emph{et~al.}
\newblock \bibinfo{title}{Amorphous {{Silicon}} with {{Extremely Low Absorption}}: {{Beating Thermal Noise}} in {{Gravitational Astronomy}}}.
\newblock \emph{\bibinfo{journal}{Phys. Rev. Lett.}} \textbf{\bibinfo{volume}{121}}, \bibinfo{pages}{191101} (\bibinfo{year}{2018}).

\bibitem{Adhikari-20-07}
\bibinfo{author}{Adhikari, R.~X.} \emph{et~al.}
\newblock \bibinfo{title}{A cryogenic silicon interferometer for gravitational-wave detection}.
\newblock \emph{\bibinfo{journal}{Class. Quantum Gravity}} \textbf{\bibinfo{volume}{37}}, \bibinfo{pages}{165003} (\bibinfo{year}{2020}).

\bibitem{Acco-96-02}
\bibinfo{author}{Acco, S.} \emph{et~al.}
\newblock \bibinfo{title}{Hydrogen solubility and network stability in amorphous silicon}.
\newblock \emph{\bibinfo{journal}{Phys. Rev. B}} \textbf{\bibinfo{volume}{53}}, \bibinfo{pages}{4415--4427} (\bibinfo{year}{1996}).

\bibitem{Kear-79-04}
\bibinfo{author}{Kear, B.~H.}, \bibinfo{author}{Breinan, E.~M.} \& \bibinfo{author}{Greenwald, L.~E.}
\newblock \bibinfo{title}{Laser glazing {\textendash} a new process for production and control of rapidly chilled metallurgical microstructures}.
\newblock \emph{\bibinfo{journal}{Metals Technology}} \textbf{\bibinfo{volume}{6}}, \bibinfo{pages}{121--129} (\bibinfo{year}{1979}).

\bibitem{Roorda-91-08}
\bibinfo{author}{Roorda, S.} \emph{et~al.}
\newblock \bibinfo{title}{Structural relaxation and defect annihilation in pure amorphous silicon}.
\newblock \emph{\bibinfo{journal}{Phys. Rev. B}} \textbf{\bibinfo{volume}{44}}, \bibinfo{pages}{3702--3725} (\bibinfo{year}{1991}).

\bibitem{Laaziri-95-11}
\bibinfo{author}{Laaziri, K.}, \bibinfo{author}{Roorda, S.} \& \bibinfo{author}{Baribeau, J.~M.}
\newblock \bibinfo{title}{Density of amorphous {Si}$_{x}${Ge}$_{1-x}$ alloys prepared by high-energy ion implantation}.
\newblock \emph{\bibinfo{journal}{J. Non-Cryst. Solids}} \textbf{\bibinfo{volume}{191}}, \bibinfo{pages}{193--199} (\bibinfo{year}{1995}).

\bibitem{Xie-13-08}
\bibinfo{author}{Xie, R.} \emph{et~al.}
\newblock \bibinfo{title}{Hyperuniformity in amorphous silicon based on the measurement of the infinite-wavelength limit of the structure factor}.
\newblock \emph{\bibinfo{journal}{Proc. Natl. Acad. Sci., U. S. A.}} \textbf{\bibinfo{volume}{110}}, \bibinfo{pages}{13250--13254} (\bibinfo{year}{2013}).

\bibitem{Fortner-89-03}
\bibinfo{author}{Fortner, J.} \& \bibinfo{author}{Lannin, J.~S.}
\newblock \bibinfo{title}{Radial distribution functions of amorphous silicon}.
\newblock \emph{\bibinfo{journal}{Phys. Rev. B}} \textbf{\bibinfo{volume}{39}}, \bibinfo{pages}{5527--5530} (\bibinfo{year}{1989}).

\bibitem{Moss-69-11}
\bibinfo{author}{Moss, S.~C.} \& \bibinfo{author}{Graczyk, J.~F.}
\newblock \bibinfo{title}{Evidence of {{Voids Within}} the {{As-Deposited Structure}} of {{Glassy Silicon}}}.
\newblock \emph{\bibinfo{journal}{Phys. Rev. Lett.}} \textbf{\bibinfo{volume}{23}}, \bibinfo{pages}{1167--1171} (\bibinfo{year}{1969}).

\bibitem{Ohdomari-81-11}
\bibinfo{author}{Ohdomari, I.} \emph{et~al.}
\newblock \bibinfo{title}{Electron paramagnetic resonance study on the annealing behavior of vacuum deposited amorphous silicon on crystalline silicon}.
\newblock \emph{\bibinfo{journal}{J. Appl. Phys.}} \textbf{\bibinfo{volume}{52}}, \bibinfo{pages}{6617--6622} (\bibinfo{year}{1981}).

\bibitem{Rosenhain-27}
\bibinfo{author}{Rosenhain, W.}
\newblock \bibinfo{title}{The structure and constitution of glass}.
\newblock \emph{\bibinfo{journal}{J. Soc. Glass Technol.}} \textbf{\bibinfo{volume}{11}}, \bibinfo{pages}{77} (\bibinfo{year}{1927}).

\bibitem{Zachariasen-32-10}
\bibinfo{author}{Zachariasen, W.~H.}
\newblock \bibinfo{title}{The {{Atomic Arrangement}} in {{Glass}}}.
\newblock \emph{\bibinfo{journal}{J. Am. Chem. Soc.}} \textbf{\bibinfo{volume}{54}}, \bibinfo{pages}{3841--3851} (\bibinfo{year}{1932}).

\bibitem{Wooten-85-04}
\bibinfo{author}{Wooten, F.}, \bibinfo{author}{Winer, K.} \& \bibinfo{author}{Weaire, D.}
\newblock \bibinfo{title}{Computer {{Generation}} of {{Structural Models}} of {{Amorphous Si}} and {{Ge}}}.
\newblock \emph{\bibinfo{journal}{Phys. Rev. Lett.}} \textbf{\bibinfo{volume}{54}}, \bibinfo{pages}{1392--1395} (\bibinfo{year}{1985}).

\bibitem{Barkema-00-08}
\bibinfo{author}{Barkema, G.~T.} \& \bibinfo{author}{Mousseau, N.}
\newblock \bibinfo{title}{High-quality continuous random networks}.
\newblock \emph{\bibinfo{journal}{Phys. Rev. B}} \textbf{\bibinfo{volume}{62}}, \bibinfo{pages}{4985--4990} (\bibinfo{year}{2000}).

\bibitem{Zallen-98-05}
\bibinfo{author}{Zallen, R.}
\newblock \bibinfo{title}{The {{Formation}} of {{Amorphous Solids}}}.
\newblock In \emph{\bibinfo{booktitle}{The {{Physics}} of {{Amorphous Solids}}}}, \bibinfo{pages}{1--32} (\bibinfo{publisher}{Wiley}, \bibinfo{address}{{Weinheim, Germany}}, \bibinfo{year}{1998}).

\bibitem{Bartok-18-12}
\bibinfo{author}{Bart{\'o}k, A.~P.}, \bibinfo{author}{Kermode, J.}, \bibinfo{author}{Bernstein, N.} \& \bibinfo{author}{Cs{\'a}nyi, G.}
\newblock \bibinfo{title}{Machine {{Learning}} a {{General-Purpose Interatomic Potential}} for {{Silicon}}}.
\newblock \emph{\bibinfo{journal}{Phys. Rev. X}} \textbf{\bibinfo{volume}{8}}, \bibinfo{pages}{041048} (\bibinfo{year}{2018}).

\bibitem{Deringer-18-06}
\bibinfo{author}{Deringer, V.~L.} \emph{et~al.}
\newblock \bibinfo{title}{Realistic {{Atomistic Structure}} of {{Amorphous Silicon}} from {{Machine-Learning-Driven Molecular Dynamics}}}.
\newblock \emph{\bibinfo{journal}{J. Phys. Chem. Lett.}} \textbf{\bibinfo{volume}{9}}, \bibinfo{pages}{2879--2885} (\bibinfo{year}{2018}).

\bibitem{Bernstein-19}
\bibinfo{author}{Bernstein, N.} \emph{et~al.}
\newblock \bibinfo{title}{Quantifying chemical structure and machine-learned atomic energies in amorphous and liquid silicon}.
\newblock \emph{\bibinfo{journal}{Angew. Chem. Int. Ed.}} \textbf{\bibinfo{volume}{58}}, \bibinfo{pages}{7057--7061} (\bibinfo{year}{2019}).

\bibitem{Baeri-82-12}
\bibinfo{author}{Baeri, P.}, \bibinfo{author}{Campisano, S.~U.}, \bibinfo{author}{Grimaldi, M.~G.} \& \bibinfo{author}{Rimini, E.}
\newblock \bibinfo{title}{Experimental investigation of the amorphous silicon melting temperature by fast heating processes}.
\newblock \emph{\bibinfo{journal}{J. Appl. Phys.}} \textbf{\bibinfo{volume}{53}}, \bibinfo{pages}{8730--8733} (\bibinfo{year}{1982}).

\bibitem{Treacy-98-07}
\bibinfo{author}{Treacy, M. M.~J.}, \bibinfo{author}{Gibson, J.~M.} \& \bibinfo{author}{Keblinski, P.~J.}
\newblock \bibinfo{title}{Paracrystallites found in evaporated amorphous tetrahedral semiconductors}.
\newblock \emph{\bibinfo{journal}{J. Non-Cryst. Solids}} \textbf{\bibinfo{volume}{231}}, \bibinfo{pages}{99--110} (\bibinfo{year}{1998}).

\bibitem{Voyles-01-11}
\bibinfo{author}{Voyles, P.~M.} \emph{et~al.}
\newblock \bibinfo{title}{Structure and physical properties of paracrystalline atomistic models of amorphous silicon}.
\newblock \emph{\bibinfo{journal}{J. Appl. Phys.}} \textbf{\bibinfo{volume}{90}}, \bibinfo{pages}{4437--4451} (\bibinfo{year}{2001}).

\bibitem{Tang-21-11}
\bibinfo{author}{Tang, H.} \emph{et~al.}
\newblock \bibinfo{title}{Synthesis of paracrystalline diamond}.
\newblock \emph{\bibinfo{journal}{Nature}} \textbf{\bibinfo{volume}{599}}, \bibinfo{pages}{605--610} (\bibinfo{year}{2021}).

\bibitem{Wright-13}
\bibinfo{author}{Wright, A.~C.} \& \bibinfo{author}{Thorpe, M.~F.}
\newblock \bibinfo{title}{Eighty years of random networks}.
\newblock \emph{\bibinfo{journal}{Phys. Status Solidi B}} \textbf{\bibinfo{volume}{250}}, \bibinfo{pages}{931--936} (\bibinfo{year}{2013}).

\bibitem{Liu-14-07}
\bibinfo{author}{Liu, X.}, \bibinfo{author}{Queen, D.~R.}, \bibinfo{author}{Metcalf, T.~H.}, \bibinfo{author}{Karel, J.~E.} \& \bibinfo{author}{Hellman, F.}
\newblock \bibinfo{title}{Hydrogen-{{Free Amorphous Silicon}} with {{No Tunneling States}}}.
\newblock \emph{\bibinfo{journal}{Phys. Rev. Lett.}} \textbf{\bibinfo{volume}{113}}, \bibinfo{pages}{025503} (\bibinfo{year}{2014}).

\bibitem{Kail-11}
\bibinfo{author}{Kail, F.} \emph{et~al.}
\newblock \bibinfo{title}{The configurational energy gap between amorphous and crystalline silicon}.
\newblock \emph{\bibinfo{journal}{Phys. Status Solidi RRL}} \textbf{\bibinfo{volume}{5}}, \bibinfo{pages}{361--363} (\bibinfo{year}{2011}).

\bibitem{Drabold-11}
\bibinfo{author}{Drabold, D.~A.}
\newblock \bibinfo{title}{Silicon: The gulf between crystalline and amorphous}.
\newblock \emph{\bibinfo{journal}{Phys. Status Solidi RRL}} \textbf{\bibinfo{volume}{5}}, \bibinfo{pages}{359--360} (\bibinfo{year}{2011}).

\bibitem{Ziman-79}
\bibinfo{author}{Ziman, J. M. J.~M.}
\newblock \emph{\bibinfo{title}{Models of disorder: the theoretical physics of homogeneously disordered systems}} (\bibinfo{publisher}{Cambridge University Press}, \bibinfo{address}{Cambridge}, \bibinfo{year}{1979}).

\bibitem{Morrow-09-22}
\bibinfo{author}{Morrow, J.~D.} \& \bibinfo{author}{Deringer, V.~L.}
\newblock \bibinfo{title}{Indirect learning and physically guided validation of interatomic potential models}.
\newblock \emph{\bibinfo{journal}{J. Chem. Phys.}} \textbf{\bibinfo{volume}{157}}, \bibinfo{pages}{104105} (\bibinfo{year}{2022}).

\bibitem{Larsen-16-05}
\bibinfo{author}{Larsen, P.~M.}, \bibinfo{author}{Schmidt, S.} \& \bibinfo{author}{Schi{\o}tz, J.}
\newblock \bibinfo{title}{Robust structural identification via polyhedral template matching}.
\newblock \emph{\bibinfo{journal}{Model. Simul. Mater. Sci. Eng.}} \textbf{\bibinfo{volume}{24}}, \bibinfo{pages}{055007} (\bibinfo{year}{2016}).

\bibitem{Laaziri-99-04}
\bibinfo{author}{Laaziri, K.} \emph{et~al.}
\newblock \bibinfo{title}{High {{Resolution Radial Distribution Function}} of {{Pure Amorphous Silicon}}}.
\newblock \emph{\bibinfo{journal}{Phys. Rev. Lett.}} \textbf{\bibinfo{volume}{82}}, \bibinfo{pages}{3460--3463} (\bibinfo{year}{1999}).

\bibitem{Laaziri-99-11}
\bibinfo{author}{Laaziri, K.} \emph{et~al.}
\newblock \bibinfo{title}{High-energy x-ray diffraction study of pure amorphous silicon}.
\newblock \emph{\bibinfo{journal}{Phys. Rev. B}} \textbf{\bibinfo{volume}{60}}, \bibinfo{pages}{13520--13533} (\bibinfo{year}{1999}).

\bibitem{Moss-70-08}
\bibinfo{author}{Moss, S.~C.} \& \bibinfo{author}{Graczyk, J.~F.}
\newblock \bibinfo{title}{Structure of amorphous silicon}.
\newblock In \emph{\bibinfo{booktitle}{Proceedings of the {{Tenth International Conference}} on the {{Physics}} of {{Semiconductors}}}}, \bibinfo{pages}{658--662} (\bibinfo{address}{Cambridge, Massachusetts, USA}, \bibinfo{year}{1970}).

\bibitem{Boissonade-74-10}
\bibinfo{author}{Boissonade, J.}, \bibinfo{author}{Gandais, M.} \& \bibinfo{author}{Theye, M.}
\newblock \bibinfo{title}{Investigations on the influence of the dihedral angle distribution on the atomic radial distribution function in amorphous {{Ge}} and {{Si}}}.
\newblock \emph{\bibinfo{journal}{J. Non-Cryst. Solids}} \textbf{\bibinfo{volume}{16}}, \bibinfo{pages}{101--109} (\bibinfo{year}{1974}).

\bibitem{Nakhmanson-01-05}
\bibinfo{author}{Nakhmanson, S.~M.}, \bibinfo{author}{Voyles, P.~M.}, \bibinfo{author}{Mousseau, N.}, \bibinfo{author}{Barkema, G.~T.} \& \bibinfo{author}{Drabold, D.~A.}
\newblock \bibinfo{title}{Realistic models of paracrystalline silicon}.
\newblock \emph{\bibinfo{journal}{Phys. Rev. B}} \textbf{\bibinfo{volume}{63}}, \bibinfo{pages}{235207} (\bibinfo{year}{2001}).

\bibitem{Holmstrom-16-04}
\bibinfo{author}{Holmstr{\"o}m, E.} \emph{et~al.}
\newblock \bibinfo{title}{Dependence of short and intermediate-range order on preparation in experimental and modeled pure a-{{Si}}}.
\newblock \emph{\bibinfo{journal}{J. Non-Cryst. Solids}} \textbf{\bibinfo{volume}{438}}, \bibinfo{pages}{26--36} (\bibinfo{year}{2016}).

\bibitem{ElMachachi-22-10}
\bibinfo{author}{El-Machachi, Z.}, \bibinfo{author}{Wilson, M.} \& \bibinfo{author}{Deringer, V.~L.}
\newblock \bibinfo{title}{Exploring the configurational space of amorphous graphene with machine-learned atomic energies}.
\newblock \emph{\bibinfo{journal}{Chem. Sci.}} \textbf{\bibinfo{volume}{13}}, \bibinfo{pages}{13720--13731} (\bibinfo{year}{2022}).

\bibitem{Toh-20-01}
\bibinfo{author}{Toh, C.-T.} \emph{et~al.}
\newblock \bibinfo{title}{Synthesis and properties of free-standing monolayer amorphous carbon}.
\newblock \emph{\bibinfo{journal}{Nature}} \textbf{\bibinfo{volume}{577}}, \bibinfo{pages}{199--203} (\bibinfo{year}{2020}).

\bibitem{Johansen-24-02}
\bibinfo{author}{Johansen, F.~L.} \emph{et~al.}
\newblock \bibinfo{title}{A {{GPU-Accelerated Open-Source Python Package}} for {{Calculating Powder Diffraction}}, {{Small-Angle-}}, and {{Total Scattering}} with the {{Debye Scattering Equation}}}.
\newblock \emph{\bibinfo{journal}{J. Open Source Softw.}} \textbf{\bibinfo{volume}{9}}, \bibinfo{pages}{6024} (\bibinfo{year}{2024}).

\bibitem{Staebler-77-08}
\bibinfo{author}{Staebler, D.~L.} \& \bibinfo{author}{Wronski, C.~R.}
\newblock \bibinfo{title}{Reversible conductivity changes in discharge-produced amorphous {{Si}}}.
\newblock \emph{\bibinfo{journal}{Appl. Phys. Lett.}} \textbf{\bibinfo{volume}{31}}, \bibinfo{pages}{292--294} (\bibinfo{year}{1977}).

\bibitem{Fedders-92-03}
\bibinfo{author}{Fedders, P.~A.}, \bibinfo{author}{Fu, Y.} \& \bibinfo{author}{Drabold, D.~A.}
\newblock \bibinfo{title}{Atomistic origins of light-induced defects in {\emph{a}}-{{Si}}}.
\newblock \emph{\bibinfo{journal}{Phys. Rev. Lett.}} \textbf{\bibinfo{volume}{68}}, \bibinfo{pages}{1888--1891} (\bibinfo{year}{1992}).

\bibitem{Fedders-96-02}
\bibinfo{author}{Fedders, P.~A.} \& \bibinfo{author}{Drabold, D.~A.}
\newblock \bibinfo{title}{Molecular-dynamics investigations of conformational fluctuations and low-energy vibrational excitations in {\emph{a}} -{{Si}}:{{H}}}.
\newblock \emph{\bibinfo{journal}{Phys. Rev. B}} \textbf{\bibinfo{volume}{53}}, \bibinfo{pages}{3841--3845} (\bibinfo{year}{1996}).

\bibitem{Erhard-24-03}
\bibinfo{author}{Erhard, L.~C.}, \bibinfo{author}{Rohrer, J.}, \bibinfo{author}{Albe, K.} \& \bibinfo{author}{Deringer, V.~L.}
\newblock \bibinfo{title}{Modelling atomic and nanoscale structure in the silicon–oxygen system through active machine learning}.
\newblock \emph{\bibinfo{journal}{Nat. Commun.}} \textbf{\bibinfo{volume}{15}}, \bibinfo{pages}{1927} (\bibinfo{year}{2024}).

\bibitem{Guttman-86-01}
\bibinfo{author}{Guttman, L.} \& \bibinfo{author}{Rahman, S.~M.}
\newblock \bibinfo{title}{Modeling a ``tunneling'' state in amorphous silicon dioxide}.
\newblock \emph{\bibinfo{journal}{Phys. Rev. B}} \textbf{\bibinfo{volume}{33}}, \bibinfo{pages}{1506--1508} (\bibinfo{year}{1986}).

\bibitem{Shapeev-16-09}
\bibinfo{author}{Shapeev, A.~V.}
\newblock \bibinfo{title}{Moment Tensor Potentials: A Class of Systematically Improvable Interatomic Potentials}.
\newblock \emph{\bibinfo{journal}{Multiscale Model. Simul.}} \textbf{\bibinfo{volume}{14}}, \bibinfo{pages}{1153--1173} (\bibinfo{year}{2016}).

\bibitem{Morrow-24-05}
\bibinfo{author}{Morrow, J.~D.} \emph{et~al.}
\newblock \bibinfo{title}{Understanding defects in amorphous silicon with million-atom simulations and machine learning}.
\newblock \emph{\bibinfo{journal}{Angew. Chem. Int. Ed.}} \textbf{\bibinfo{volume}{63}}, \bibinfo{pages}{e202403842} (\bibinfo{year}{2024}).

\bibitem{Stukowski-10-01}
\bibinfo{author}{Stukowski, A.}
\newblock \bibinfo{title}{{Visualization and analysis of atomistic simulation data with OVITO-the Open Visualization Tool}}.
\newblock \emph{\bibinfo{journal}{{Model. Simul. Mater. Sci. Eng.}}} \textbf{\bibinfo{volume}{{18}}}, \bibinfo{pages}{015012} (\bibinfo{year}{{2010}}).

\bibitem{Bartok-13-05}
\bibinfo{author}{Bart{\'o}k, A.~P.}, \bibinfo{author}{Kondor, R.} \& \bibinfo{author}{Cs{\'a}nyi, G.}
\newblock \bibinfo{title}{On representing chemical environments}.
\newblock \emph{\bibinfo{journal}{Phys. Rev. B}} \textbf{\bibinfo{volume}{87}}, \bibinfo{pages}{184115} (\bibinfo{year}{2013}).

\bibitem{Honeycutt-87-09}
\bibinfo{author}{Honeycutt, J. D.} \& \bibinfo{author}{Andersen, H. C.}
\newblock \bibinfo{title}{Molecular dynamics study of melting and freezing of small Lennard-Jones clusters}.
\newblock \emph{\bibinfo{journal}{J. Phys. Chem.}} \textbf{\bibinfo{volume}{91}}, \bibinfo{pages}{4950--4963} (\bibinfo{year}{1987}).

\bibitem{Maras-16-08}
\bibinfo{author}{Maras, E.}, \bibinfo{author}{Trushin, O.}, \bibinfo{author}{Stukowski, A.}, \bibinfo{author}{{Ala-Nissila}, T.} \& \bibinfo{author}{J{\'o}nsson, H.}
\newblock \bibinfo{title}{Global transition path search for dislocation formation in {{Ge}} on {{Si}}(001)}.
\newblock \emph{\bibinfo{journal}{Comput. Phys. Commun.}} \textbf{\bibinfo{volume}{205}}, \bibinfo{pages}{13--21} (\bibinfo{year}{2016}).

\bibitem{Behler-07-04}
\bibinfo{author}{Behler, J.} \& \bibinfo{author}{Parrinello, M.}
\newblock \bibinfo{title}{Generalized neural-network representation of high-dimensional potential-energy surfaces}.
\newblock \emph{\bibinfo{journal}{Phys. Rev. Lett.}} \textbf{\bibinfo{volume}{98}}, \bibinfo{pages}{146401} (\bibinfo{year}{2007}).

\bibitem{Bartok-10-04}
\bibinfo{author}{Bart\'ok, A.~P.}, \bibinfo{author}{Payne, M.~C.}, \bibinfo{author}{Kondor, R.} \& \bibinfo{author}{Cs\'anyi, G.}
\newblock \bibinfo{title}{Gaussian approximation potentials: {The} accuracy of quantum mechanics, without the electrons}.
\newblock \emph{\bibinfo{journal}{Phys. Rev. Lett.}} \textbf{\bibinfo{volume}{104}}, \bibinfo{pages}{136403} (\bibinfo{year}{2010}).

\bibitem{Wang-23-02}
\bibinfo{author}{Wang, S.}, \bibinfo{author}{Liu, Y.} \& \bibinfo{author}{Mo, Y.}
\newblock \bibinfo{title}{Frustration in {Super}-{Ionic} {Conductors} {Unraveled} by the {Density} of {Atomistic} {States}}.
\newblock \emph{\bibinfo{journal}{Angew. Chem. Int. Ed.}} \textbf{\bibinfo{volume}{62}}, \bibinfo{pages}{e202215544} (\bibinfo{year}{2023}).

\end{thebibliography}
\end{document}

\title{\Large {\bf Supplementary Material for}\\ {`Signatures of paracrystallinity in amorphous silicon'}}

\author[1]{Louise A. M. Rosset}
\author[2]{David A. Drabold}
\author[1]{Volker L. Deringer\thanks{volker.deringer@chem.ox.ac.uk}}

\affil[1]{Department of Chemistry, University of Oxford, Oxford, UK}
\affil[2]{Department of Physics and Astronomy, Ohio University, Athens, OH, USA}

\date{}

\maketitle

\setstretch{1.5}
\setcounter{suppfigure}{1}

\newpage
\section{Methods}
\subsection{Structure generation}
\subsubsection{Dataset}
\begin{figure}
    \centering
    \includegraphics{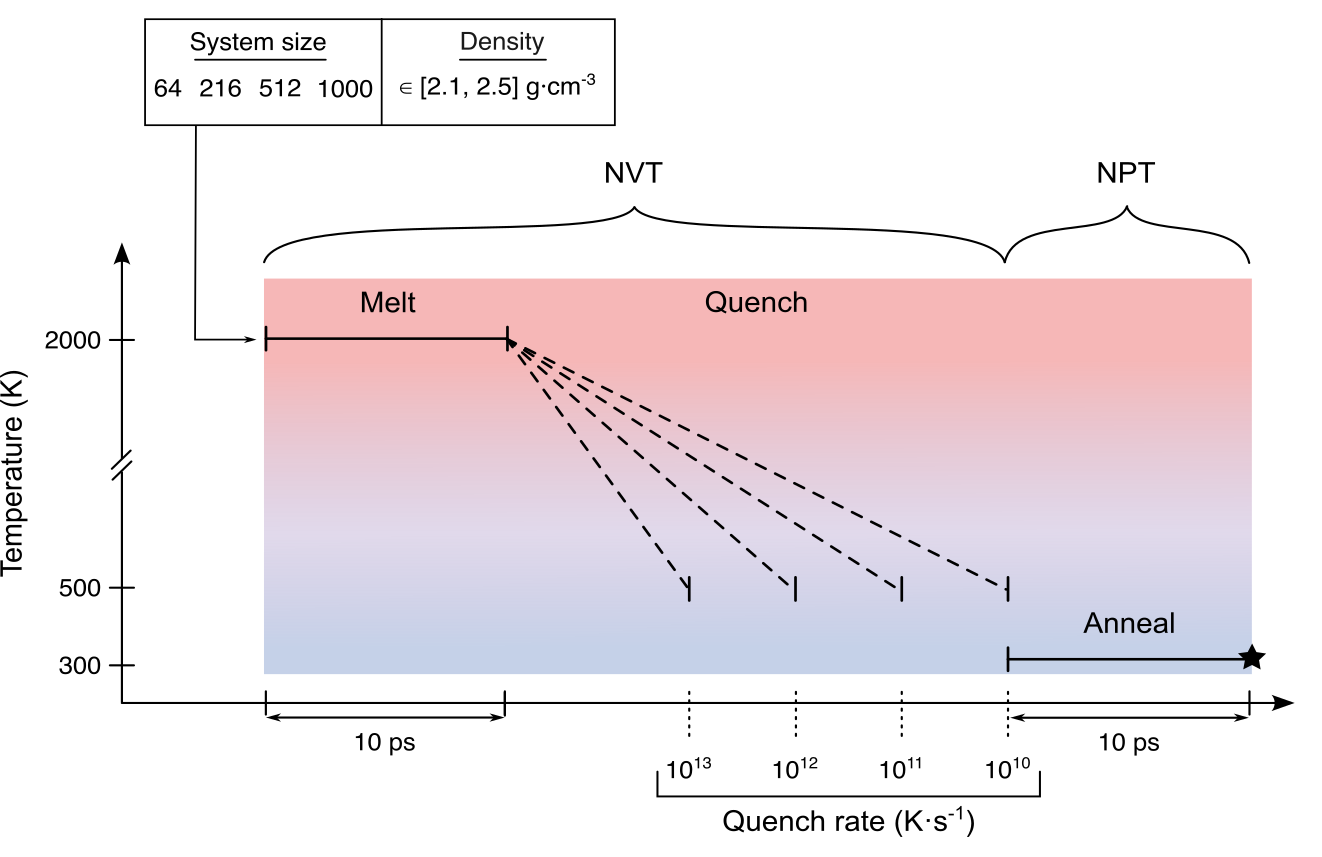}
    \caption{Schematic overview of the protocol used for the molecular-dynamics simulations reported in the present work. The last frame from the simulation is indicated by a star; this frame is added to our dataset.}
\label{fig:DBprotocol}
\refstepcounter{suppfigure}
\end{figure}

To build our dataset of silicon configurations, we follow the protocol depicted in Fig.~\ref{fig:DBprotocol} using LAMMPS. \cite{Thompson-22-02} We start with randomized structures of varying system size (64, 216, 512, and 1,000 atoms) and uniformly sample densities between 2.1 and 2.5 g cm$^{-3}$ with a step size of 0.002 g cm$^{-3}$, chosen to provide a range of underdense and overdense a-Si structures. These structures are melted at 2,000 K for 10 ps, then quenched to 500 K at various quench rates from 10$^{13}$ K/s to 10$^{10}$ K/s in the canonical ensemble (NVT). Finally they are annealed for 10 ps at 300 K in the isothermal–isobaric ensemble (NPT). The final structure is added to the dataset. We employ a Nosé--Hoover thermostat for the NVT simulations and a Nosé--Hoover thermostat and barostat for the NPT simulations. These simulations are run with a timestep of 1 fs. 

The melt-quench simulations of 1,000 atoms at rates of 10$^{10}$ K/s for structures with densities $\rho >2.29$ g cm$^{-3}$ were omitted as they were computationally expensive and systematically resulted in polycrystalline structures with high counts of diamond-like environments, not highly relevant to our study of paracrystalline silicon, and we already had many such structures.

\subsubsection{Large-scale models}
The 100,000-atom structures with 0.8$\%$ and 62.3$\%$  of diamond-like environments were generated by quenching randomized cells of density $\rho=2.252$ \si{g.cm^{-3}} at rates of 10$^{11}$ and 10$^{10}$ K/s respectively, following the same protocol as laid out in Fig.~\ref{fig:DBprotocol}. We substitute the anneal treatment for an NPT annealing run at 300 K for 50 ps with the Si-GAP-18 potential. \cite{Bartok-18-12}

\subsection{Choice of potential}
\begin{figure}
    \centering
    \includegraphics{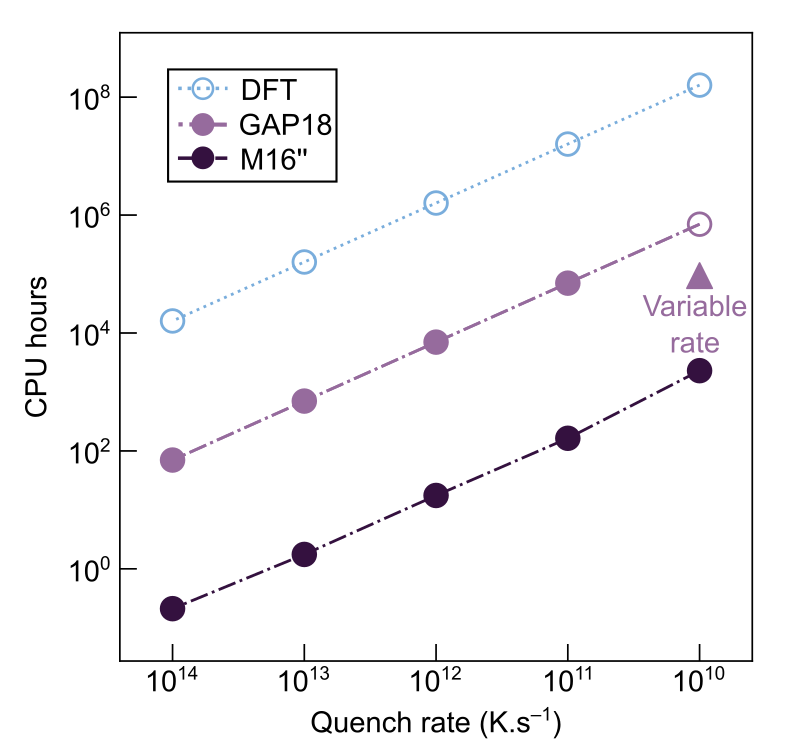}
    \caption{Comparison of the computational cost of quench simulations of a 512-atom system at different quench rates using DFT-, Si-GAP-18- (`GAP18') and M$_{16}''$-driven MD, drawn similar to Ref.~\citenum{Deringer-18-06}. DFT computational time estimates are taken from Ref.~\citenum{Deringer-18-06}.}
    \label{fig:speed-comp}
    \refstepcounter{suppfigure}
\end{figure}

We use the M$_{16}''$ `student' potential of Ref.~\citenum{Morrow-09-22} to drive our MD simulations. This potential, fitted using the Moment Tensor Potential (MTP) framework, \cite{Shapeev-16-09, Novikov-20-12} is two orders of magnitude faster than its `teacher' counterpart, Si-GAP-18, \cite{Bartok-18-12} as seen in Fig.~\ref{fig:speed-comp}. The teacher--student approach therefore allows us to use M$_{16}''$ to simulate quench rates as slow as 10$^{10}$ K/s, which are not accessible for slow quenches of large system sizes using the teacher model.

\begin{figure}
    \centering
    \includegraphics{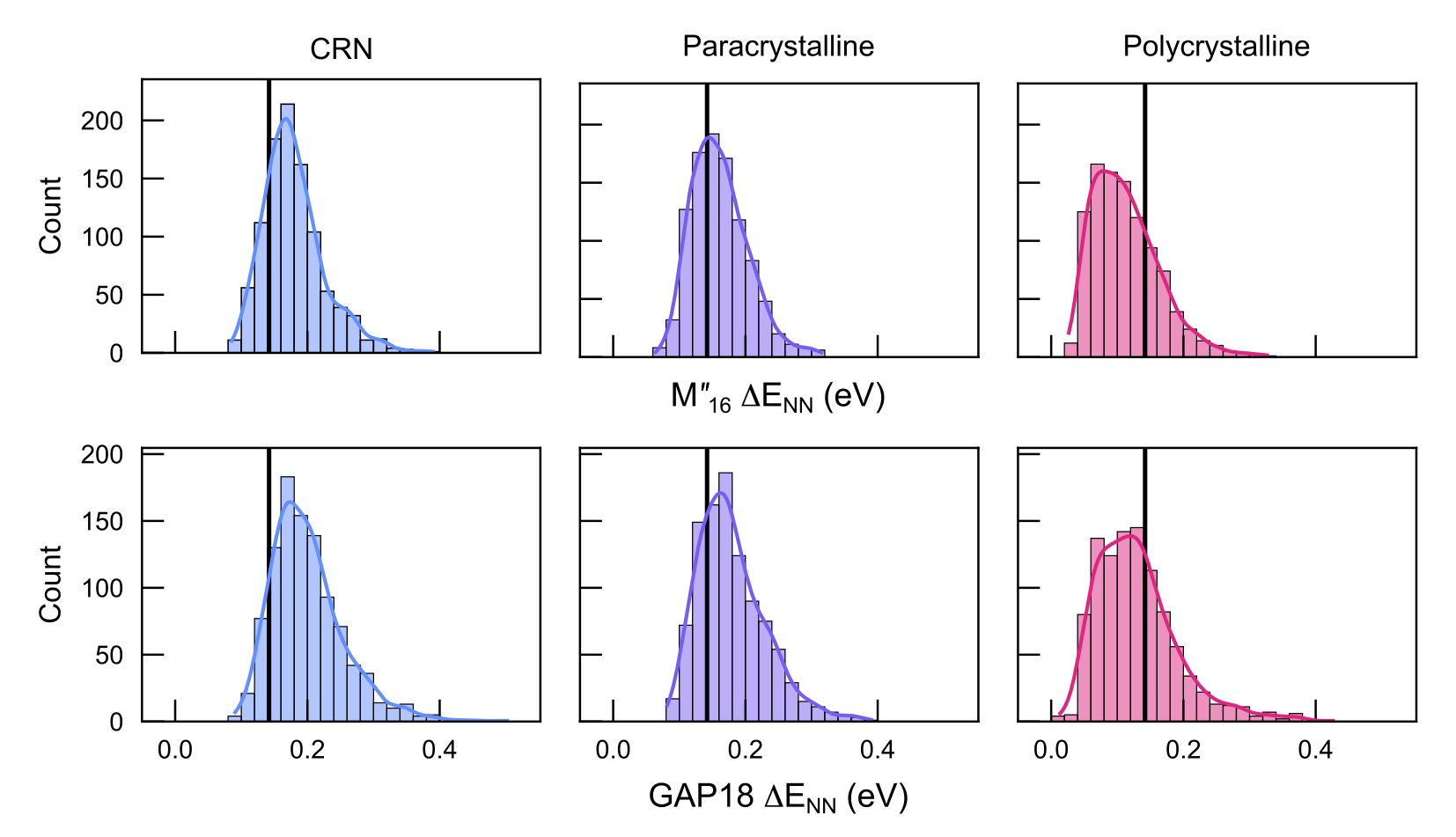}
    \caption{Comparison of the local energy prediction by the M$_{16}''$ ({\em top row}) and Si-GAP-18 ({\em bottom row}) potentials for the structures analyzed in Fig.~3.}
    \label{fig:pred-comp}
    \refstepcounter{suppfigure}
\end{figure}

To verify the energy predictions by the M$_{16}''$ potential, we compare the local-energy predictions, averaged over nearest neighbors, by M$_{16}''$ to those by Si-GAP-18. \cite{Bartok-18-12} The predictions and overall trends computed with M$_{16}''$, shown in Fig.~\ref{fig:pred-comp}, are in line with the results of Si-GAP-18.

\subsection{Structural analysis}
\subsubsection{Polyhedral Template Matching}
We use Polyhedral Template Matching (PTM) \cite{Larsen-16-05} as implemented in OVITO \cite{Stukowski-10-01} to identify the local crystalline structure of atomic environments. This classification method matches the atomic neighborhood of an atom to templates of different crystalline structures using convex hulls, and assigns crystalline structure types within a deviation metric (RMSD cutoff), or the type `other' beyond the cutoff. We choose a RMSD cutoff of 0.1 for our analysis at 300 K, in accordance with Ref.~\citenum{Larsen-16-05}.

Both cubic diamond ({\bf dia}) and hexagonal diamond  ({\bf lon}) templates contain four first-neighbor atoms and eight second-neighbor atoms. The structure types differ by the conformation of the 6-membered rings---which is `chair' for {\bf dia} and a combination of `chair' and `boat' for {\bf lon}. The majority of neighborhoods identified as {\bf lon}-like environments are located at grain boundaries, grain--matrix boundaries, or are stacking defects. This assignment to {\bf lon} rather than {\bf dia} could be an artefact of the template method. 

We then assign each structure to a category depending on the proportion of locally `crystal-like' atoms identified by PTM, as described in the Methods section of the main text. We choose a threshold of 15$\%$ as a limit between the `paracrystalline' and `polycrystalline' categories as our dataset has a deficit in structures with local crystallinity between 15 and 40 $\%$. Furthermore, we see clustering of the diamond-like environments beyond 15$\%$, leading to structures with true grains rather than isolated crystalline environments.

\subsubsection{Common Neighbor Analysis}
We use Common Neighbor Analysis (CNA) to identify neighborhoods around crystalline environments. \cite{Honeycutt-87-09, Stukowski-12-05, Maras-16-08} While CNA is less reliable than PTM \cite{Larsen-16-05} as it uses Euclidean distances to match environments to structure types, it provides information about the neighborhoods around identified crystalline-like atomic environments.

\subsubsection{Medium-range order}
For the analysis of the medium-range order characteristics in Fig.~2., we chose four paracrystalline structures of increasing paracrystallinity from the set of structures of 1,000 atoms and generated by melt-quenching at a rate of 10$^{11}$ K/s for consistency in the structural quality of these models. Similarly, the CRN chosen for comparison is also from a 10$^{11}$ K/s quench.

\subsubsection{Structure factor}
\begin{figure}
    \centering
    \includegraphics{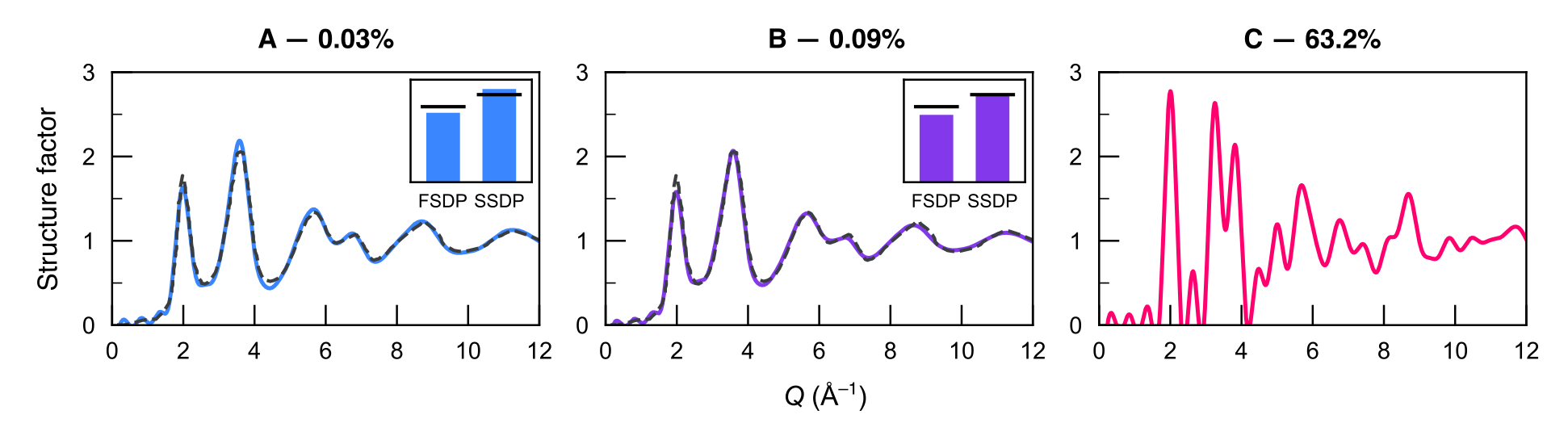}
    \caption{Computed structure factor, as obtained via the Fourier transform of the radial distribution function, for each structure presented in Fig.~4a of the main text. Black dashed lines indicate the experimental data from Ref.~\citenum{Laaziri-99-04}. The insets show the agreement of the first and second sharp diffraction peaks with the experimental data (black lines).}
    \label{fig:sq}
    \refstepcounter{suppfigure}
\end{figure}

We computed the structure factor in an alternative way, via the Fourier transform of the radial distribution function, as shown in Fig.~\ref{fig:sq}.

\newpage
\section{Supplementary results}
\subsection{Dataset}
\begin{figure}
    \centering
    \includegraphics{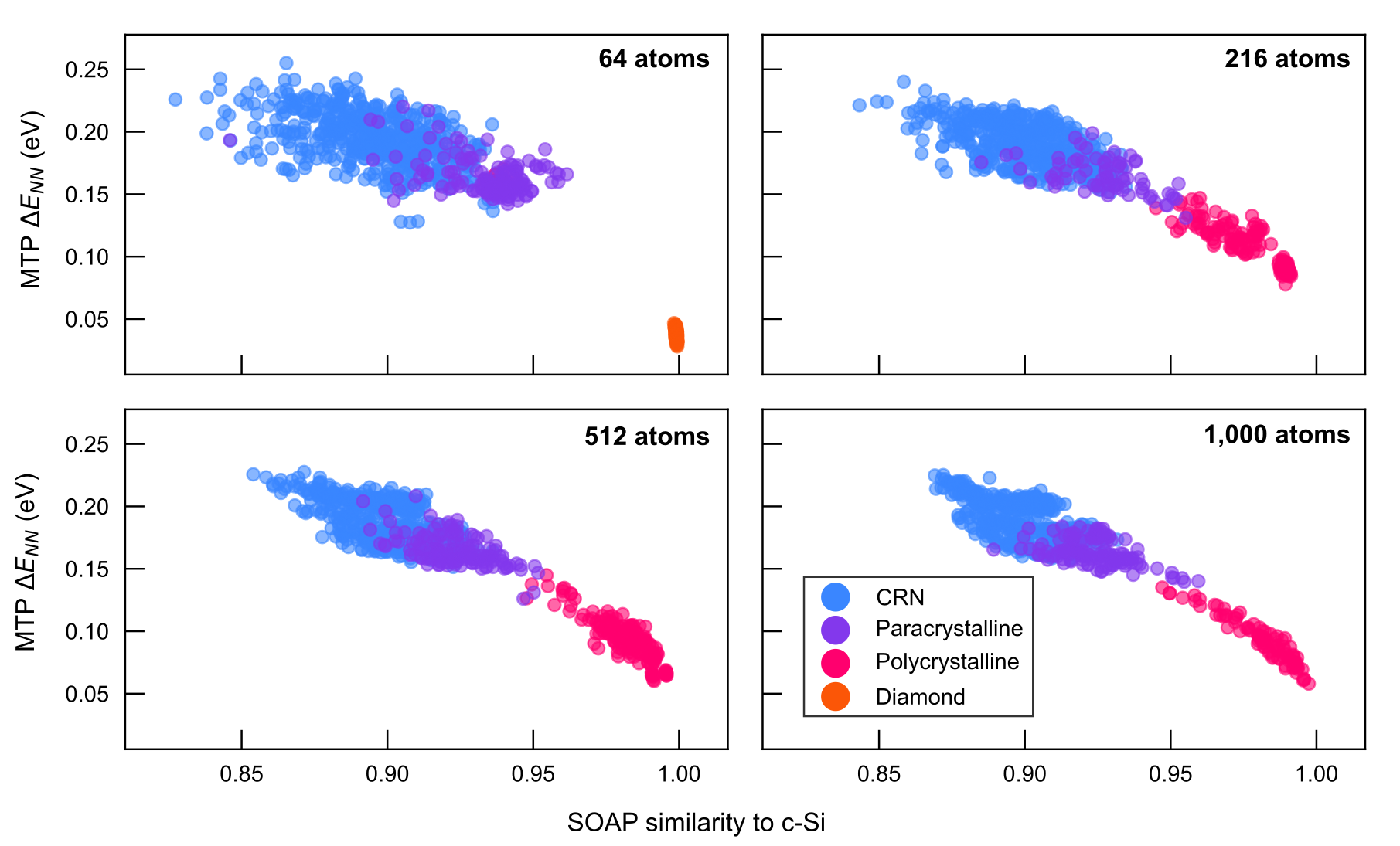}
    \caption{Dataset plotted by cell size contribution with 801 structures of 64 atoms, 782 structures of 216 atoms, 797 structures of 512 atoms and 689 structures of 1,000 atoms.}
    \label{fig:f1-byatoms}
    \refstepcounter{suppfigure}
\end{figure}

To visualize the effect of different simulation-cell sizes in the dataset, we plot the contribution of each cell size to the total dataset in Fig.~\ref{fig:f1-byatoms}. The 64-atom cells are the only ones that fully crystallize (orange markers at a SOAP similarity close to 1), and provide more strained CRN environments (light blue markers). These CRN environments make up the first of the two CRN peaks shown in the histogram in Fig.~1b of the main text. The 216-, 512-, and 1,000-atom cells all have similar contributions to the dataset in terms of the structural diversity covered.

\begin{figure}
    \centering
    \includegraphics{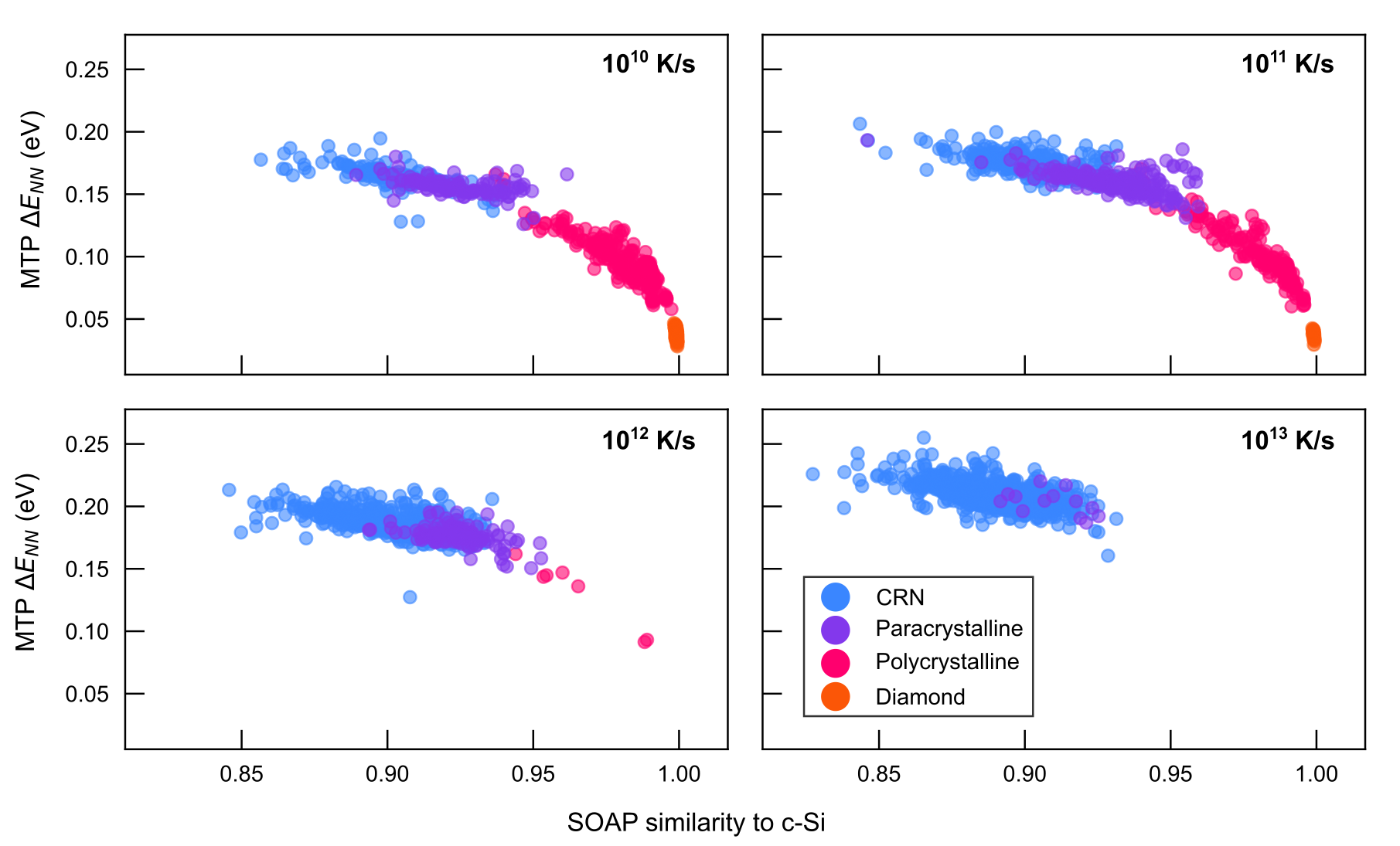}
    \caption{As Fig.~\ref{fig:f1-byatoms}, but now showing the results separately according to the quench rate.}
    \label{fig:f1-byrates}
    \refstepcounter{suppfigure}
\end{figure}

In Fig.~\ref{fig:f1-byrates}, we characterize the dataset using separate plots for each quench rate. Partial and full crystallization occurs at quench rates of 10$^{10}$ and 10$^{11}$ K/s. Even at the same quench rate, structures of different densities can amorphize or crystallize. Fast quenching at 10$^{13}$ K/s did not lead to crystallization.

\newpage
\subsection{Continuous Random Networks}
\begin{figure}
    \centering
    \includegraphics{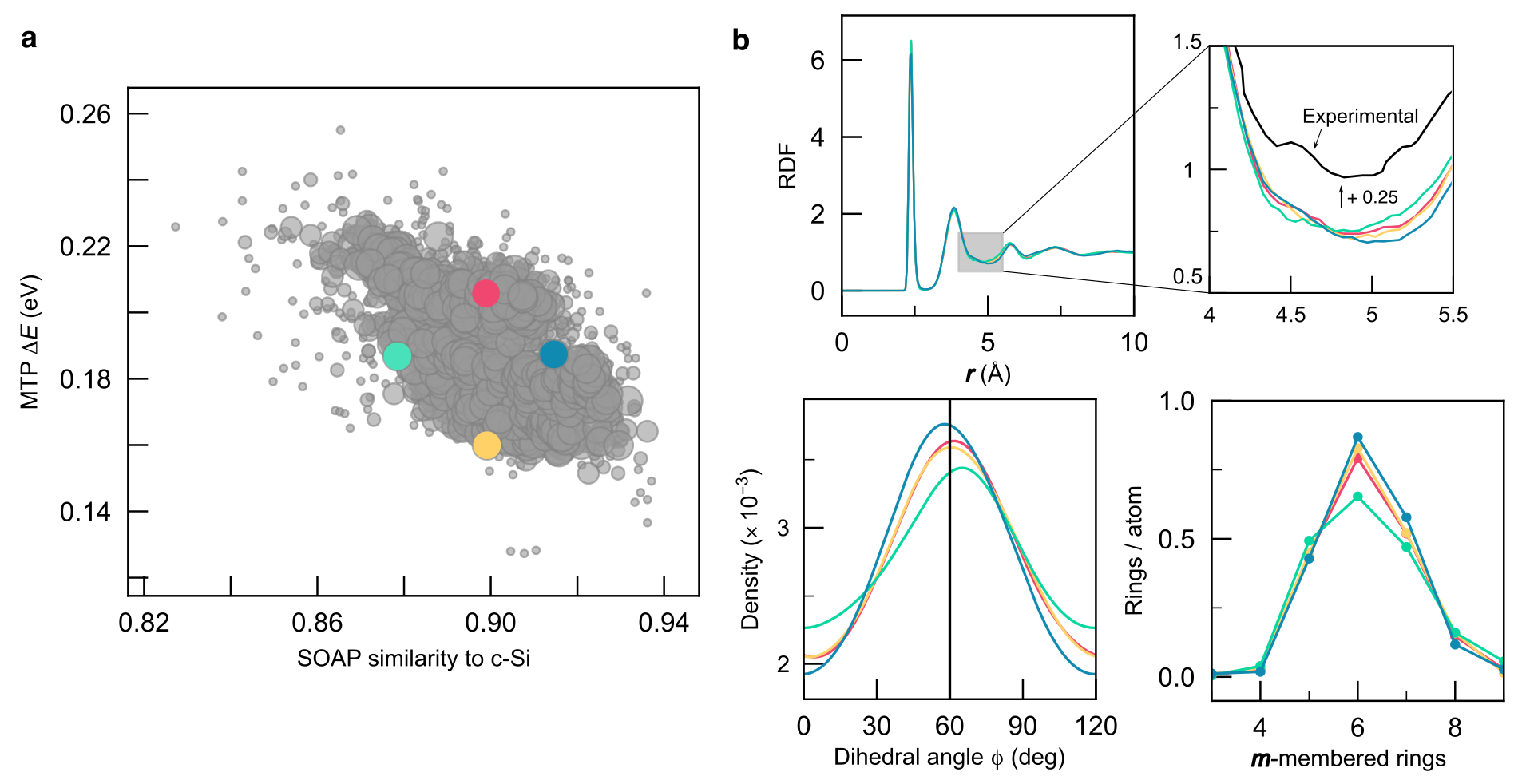}
    \caption{Comparison of CRN models of 1,000 atoms. (a) Map of similarity to diamond-type Si against the predicted excess energy for CRN structures only, where four selected structures have been colored. (b) Characteristics of medium-range order for the four selected structures.}
    \label{fig:mro-crn}
    \refstepcounter{suppfigure}
\end{figure}

We compare four structures of 1,000 atoms from the `CRN' category to investigate the structural diversity within the CRNs in our dataset. In particular, we choose two CRN structures of essentially the same SOAP similarity ($=0.899$) in red and yellow in Fig.~\ref{fig:mro-crn} a), and two structures of the essentially same energy ($\Delta E=0.205$ eV) in green and blue, and investigate their medium-range order characteristics in Fig.~\ref{fig:mro-crn}b. Importantly, we could only choose structures generated from melt-quench simulations at the same quench rate to ensure that there is no influence due to the speed of the quench. Here we choose a quench rate of 10$^{13}$ K/s, as it provides the largest diversity of environments.

The radial distribution functions of the four CRNs are very similar and show no enhancement between the second and third peaks. They show some variation in both the dihedral angle and ring distributions, but within a smaller range than that of the paracrystalline studied in Fig.~2. The structure indicated in green is noticeably different from the others, which is likely due to its low density of $\rho=2.152$ g cm$^{-3}$ compared to the others. Some of the variation in the medium-range order can also be attributed to the fast quench rate.

\newpage
\subsection{Range of paracrystallinity}
\subsubsection{1,000-atom structures}
\begin{figure}
    \centering
    \includegraphics{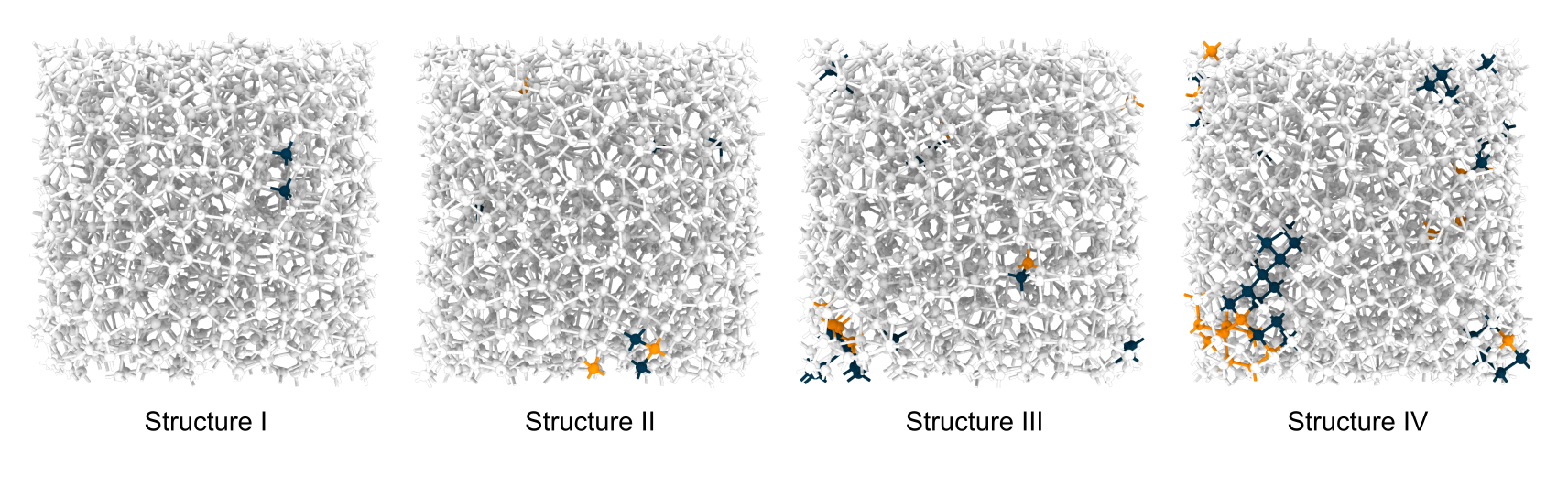}
    \caption{Visualization of the four paracrystalline structures analyzed in Fig.~2, using \textsc{OVITO}. \cite{Stukowski-10-01}}
    \label{fig:1krenders}
    \refstepcounter{suppfigure}
\end{figure}

\begin{figure}
    \centering
    \includegraphics{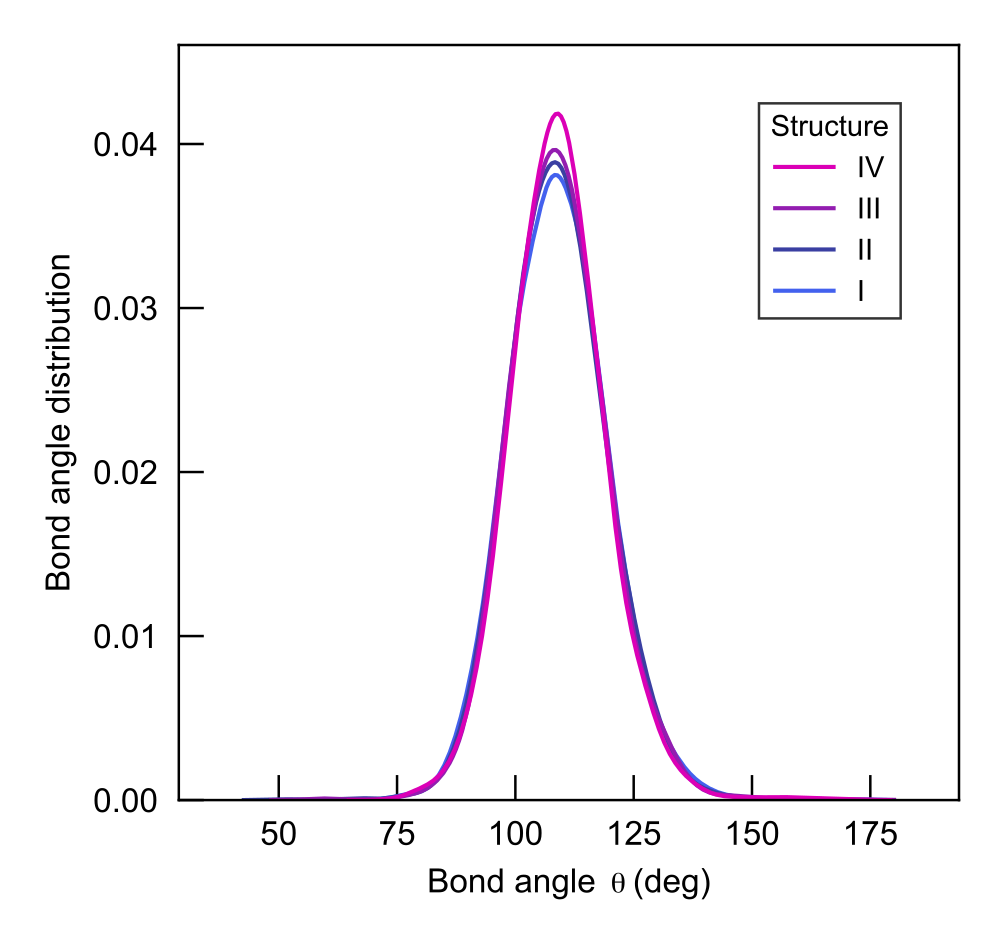}
    \caption{Bond-angle distribution of the four paracrystalline structures analyzed in Fig.~2.}
    \label{fig:SRO}
    \refstepcounter{suppfigure}
\end{figure}

The structures investigated in Fig.~2 and depicted in Fig.~\ref{fig:1krenders} present little variation in their bond-angle distribution and short-range order, as shown in Fig.~\ref{fig:SRO}.

\newpage
\subsubsection{Other system sizes}
\begin{figure}
    \centering
    \includegraphics{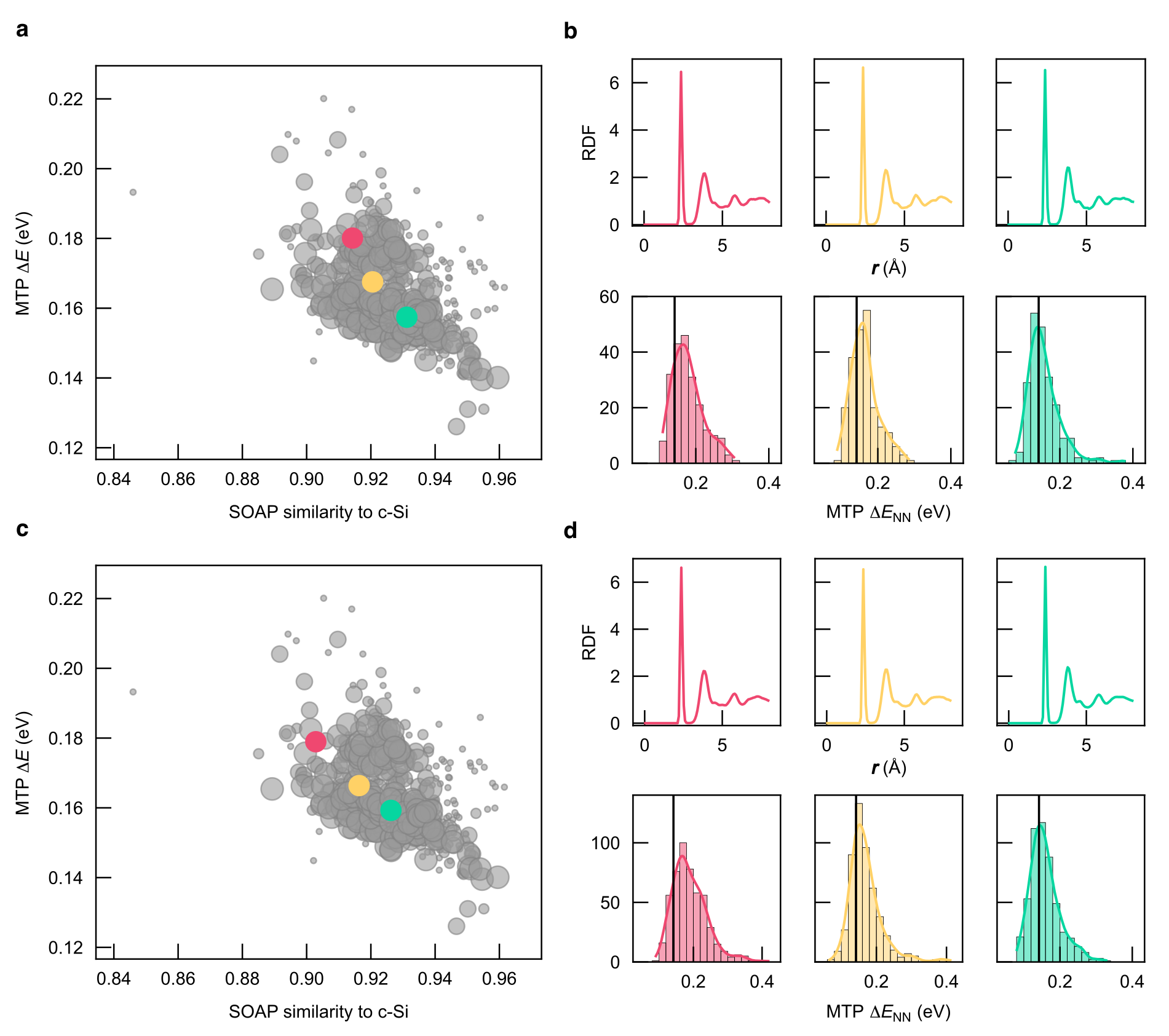}
    \caption{A study of paracrystalline structures of 216 atoms (top) and 512 atoms (bottom). Three paracrystalline structures of increasing paracrystallinity are chosen from the set of paracrystalline structures, highlighted in (a). Their radial distribution function and energetics are presented in (b). The same protocol is applied for structures of 512 atoms in (c) and (d).}
    \label{fig:para-other}
    \refstepcounter{suppfigure}
\end{figure}

We replicate the study of paracrystallinity on the other system sizes in our dataset, viz.\ 216 and 512 atoms, in Fig.~\ref{fig:para-other}. We omit the structures of 64 atoms per cell, as they are too small for meaningful radial distribution functions.

We see that these smaller structures reproduce our findings, that paracrystallinity influences medium-range order and tends to lower the total energy.

\newpage
\subsection{Paracrystalline grain stability}
To evaluate the stability of our paracrystalline structures and to determine whether the local ordering could be the onset of crystallization, we carried out high-temperature annealing simulations for the structures presented in Fig.~4.
The structures were held at temperatures ranging from 1,000 to 1,500 K for 1 ns in the NPT ensemble.
This protocol was repeated five times at each temperature. The evolution of the percentage of diamond-like environments, both \textbf{dia} and \textbf{lon}, is plotted as a function of the progress of the simulation.
At temperatures below 1,200 K, diffusion is limited and the count of diamond environments is stable. At 1,200 K, sufficient energy is supplied for diffusion and both the CRN and paracrystalline structures see the nucleation and growth of a crystalline grain, although the onset of crystallization of the grain is earlier in the case of the paracrystalline model than it is for the CRN. The polycrystalline structure is stable at this temperature. For all annealing simulations at $T>1,300$ K, the structures melt and the count of diamond-like environments drops to zero.

\begin{figure}
    \centering
    \includegraphics{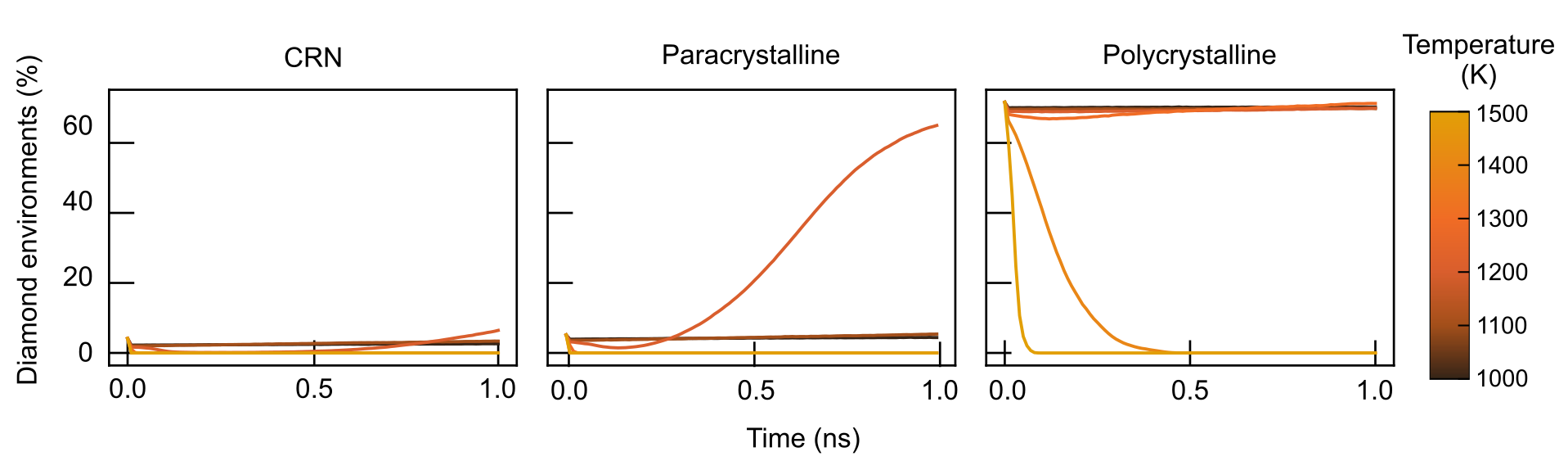}
    \caption{Evolution of the count of diamond-like environments during high-temperature annealing simulations for the structures presented in Fig.~4 of the main text.}
    \label{fig:anneals}
    \refstepcounter{suppfigure}
\end{figure}